\documentstyle[prb,twocolumn,eqsecnum,aps]{revtex}

\begin{document}
\draft
\narrowtext

\title{Electronic Excitations and Stability of the Ground State \\
of $C_{60}$ Molecules}

\author{F. Bechstedt,$^{1,2}$ M. Fiedler,$^2$ and L.J. Sham$^1$}

\address{$^1$Department of Physics, University of California San Diego,
La Jolla, California 92093 -0319 \\
$^2$Institut f\"ur Festk\"orpertheorie und Theoretische Optik,
Friedrich-Schiller-Universit\"at, 07743 Jena, Germany}

\date{\today}
\maketitle
\begin{abstract}
A model study of the singlet excitons in the $C_{60}$ molecule with emphasis on
the Coulomb interaction between excited electron and hole leads to a physical
understanding of the interaction effects on the absorption spectra and to a new
identification of the forbidden excitons in the third-harmonic generation spectra.
These conclusions may be tested experimentally on the model predictions related
to the optical Kerr effect.  The model shows that, with sufficiently strong
interatomic than onsite interaction, a $T_{2G}$ exciton could have very low
energy or become unstable against the closed-shell ground state.  
Properties of these interesting cases beyond the $C_{60}$ are briefly examined.
\end{abstract}

\section{Introduction}

Since the discovery of $C_{60}$ \cite{HWK-85}, several experimental
and theoretical studies have suggested that undoped and doped solid $C_{60}$
are strongly correlated electron systems. The observational basis includes the
unusually high superconducting transition temperature of the alkali-metal-doped
fullerites \cite{SHG-91}, the existence of soft ferromagnetism
\cite{FM-91}, and the strong Coulomb interaction effects in the Auger, direct
and inverse photoemission spectra \cite{RWL-92}.  The theoretical motivation is
based on the narrow bandwidth compared to the strong intra-molecular
interaction.  This naturally leads to theories based on the Hubbard model with
onsite interaction on every carbon atom \cite{RWL-92,BF-92,Dong-96}.  Strong
correlation is used to find a mechanism for superconductivity
\cite{Lammert-95}.  Parity doublets of the lowest unoccupied molecular orbitals
(LUMO's) of the $C_{60}$ molecule are used for the pairing mechanism
\cite{RF-92}.  Coulomb interaction of the electrons within a single $C_{60}$
molecule has also taken into account through configuration interaction by
several quantum-chemistry calculations
\cite{SL-92,ILLU-89,RDB-91,FN-92}.  On the other hand, quasiparticle
correction \cite{FB-92,ELS-93} to the local density approximation
\cite{SS-91,Ching-91} leads to the conclusion that, while the correlation
effect is large, the electronic structure in $C_{60}$ is nonetheless that of a
standard band insulator \cite{ELS-93}.  Similarly, the superconducting
transition temperature of the alkali-doped $C_{60}$ has been explained by the
usual phonon mechanism \cite{Varma,Schluter}.

A common feature among the theories mentioned above is the assumption of the
closed-shell ground state for $C_{60}$ with the set of highest occupied
molecular orbitals (HOMO's) of symmetry $h_u$ and $h_g$ and the set of LUMO's of
symmetry $t_{1u}$ and $t_{1g}$.  We ask the question whether such a ground
state is stable against the excitation of an electron from a HOMO to a LUMO.  If
not, the new ground state would lead to low-lying excited states of a quite
different nature.  In this paper, we use the $C_{60}$ molecule as a paradigm for
molecular solids to investigate the stability of the closed shell ground state.
By using a simple model, we hope to understand the factors governing such
instability and to explore consequences, such as in nonlinear optical properties. 
Even if the excited states are only low in energy without causing any instability,
they could play an interesting role in some properties, especially
superconductivity.  The method of this study may be applied to molecular solids
and quantum dots besides $C_{60}$.

First, we address the energy ordering of the closed-shell state and the
one electron-hole pair excited states.  Shirley et al.\ \cite{Shirley} used a
molecular orbital model to make a comprehensive study of the exciton energy
spectrum in solid $C_{60}$.  We adopt a similar approach but use further
simplifications.  We restrict our attention to a single
$C_{60}$ molecule since in the solid the weak overlapping between the
molecules \cite{RFC-91} would not qualitatively affect our results.  We use a
nearest-neighbor tight-binding model for the one-electron $\pi$ orbitals
\cite{RF-92,RCH-86,DAB-73} with the help of symmetry considerations
\cite{YD-92}.  The essential results of this model are given in
Sec.~\ref{sec2}.  In Sec.~\ref{sec3}, the energies of the electron-hole pair
states relative to the closed-shell state are determined in terms of the
single-particle energies and the Coulomb interaction.  The interaction terms
include direct electron-hole attraction and the exchange counterpart
\cite{ShamRice}, interaction on the same carbon site as well as between any
pairs of carbon sites in the same $C_{60}$ molecule.  The dependence of the
pair-state energies on three model parameters, the nearest-neighbor hopping
energy $V$, the intrasite interaction $U$ (including both the direct and exchange
contributions) and the typical long-range interaction term $e^2/(\epsilon R_0)$
taking into account the intersite screening effect $\epsilon$ and setting the
distance scale at the radius of the buckyball, $R_0$ ($\approx 3.5$~\AA) is
discussed.  We find that a scenario of either the closed-shell state or an
electron-hole pair state being the lowest energy state is possible for reasonable
values of the three parameters.  The consequences of both scenarios on the
linear and nonlinear optical properties are then considered in Sec.~\ref{sec4}.
Our calculations of the linear optical spectra are compared with experiment and
our calculations of the nonlinear properties are used to suggest measurements
which will clarify the situation.  Sec.~\ref{sec5} discusses the results of our
work with regard to the nature of the ground state of $C_{60}$.  Our tentative
conclusion is that the comparison of the theoretical and experimental linear
optical spectra favors the closed-shell state as the ground state.  However,
there are some unsatisfactory features.  We believe that the electro-optic and
nonlinear optical measurements can clarify the situation.  We speculate on the
possibility of molecules with an excitonic state as the ground state.

\section{Single-Particle Excitations}
\label{sec2}

Since one objective of this work is the study of the stability of the
closed-shell ground state against the low-lying excited electron-hole pair
states, the first order of business is to construct the most relevant
one-electron orbitals in $C_{60}$, namely the HOMO's and LUMO's.  Their
approximation by $\pi$ orbitals seems to be well established for
low-energy excitations. For instance, the weights of the radial orbital
for the $h_u$ and $t_{1u}$ states were about 98\% and 95\%, respectively,
according to Laouini et al. \cite{NL-95}. In the discussion of the
electron-hole interaction the small
$\sigma$-contributions may be negligible.  The
tight-binding Hamiltonian with nearest-neighbor hopping may be built from the
molecular $\pi$ orbitals \cite{YD-92}:
\begin{equation}
\phi_{lmp}({\bf x})=N_l\sum_g e_{lmp}({\bf g})\chi\left({\bf x}-R_0{\bf
g}\right)
\end{equation}
where $\chi({\bf x}-R_0{\bf g})$ is the component of the $p$ wave function
centered around the atomic sites $R_0{\bf g}$ pointing along the radial
direction.  We have neglected the difference in bond lengths of the
inequivalent bonds.  $N_l$ gives the normalization of the molecule state.  The
quantum number $l$ is used to index the irreducible representations
\cite{MH-62} of the icosahedral group $I_h$, the symmetry group of the 
buckyball, 
namely $a$, $t_1$, $h$, $t_2$, and $g$, with degeneracies 1, 2, 5, 3, and
4, respectively.  The quantum number $m$ runs over the degenerate states of
each irreducible representation.  The quantum number $p$ denotes the parity of
the state. Its introduction indicates that the full symmetry group of
$C_{60}$ is $I_h\times Z_2$, where $Z_2$ is the two-element group consisting
of the inversion operator and the identity.
The coefficients for different sites may be
related to each other by \cite{RF-92}
\begin{equation}
e_{lpm}({\bf g})=\sum^{+l}_{m'=-l}D^{l^{*}}_{mm'}(\omega_g)e_{lpm'}({\bf e}),
\label{2.2}
\end{equation}
where $\omega_g$ is the rotation bringing the radial vector from atomic site
${\bf e}$ (see Fig.~1, not to be confused with the coefficients 
$e_{lmp}({\bf g})$) to site ${\bf g}$. 
For the regular three- and five-dimensional representations $(l=1,2)$ under
consideration
the $(2l+1)\times(2l+1)$ matrices $\hat{D}^l$ are simply the standard
transformation matrices in a rigid body, the so-called Wigner-$D$ functions
\cite{DAV-88}. 

The irreducible representations $D^{l}_{mm'}(\omega_g)$ of the coefficients
$e_{lmp}({\bf g})$ reduce the nearest-neighbor Hamiltonian in units of the
hopping parameter $-V$ to a set of Hamiltonians given by \cite{RF-92}:
\begin{equation}
h^l_{mm'}=\sum^3_{i=1}D^{l^{*}}_{mm'}\left( \omega_{f_i}\right),
\end{equation}
where the $i$-sum only runs over the three nearest neighbors
${\bf f}_i$ of the site ${\bf e}$.  The spin degrees of freedom are understood.
For readers interested in generating the irreducible representations
\cite{DAV-88}, we record the coordinates
${\bf e}=\frac{1}{3}\frac{R}{R_0} (\sin(2\Theta_0),0,2+\cos(2\Theta_0))$, where
the angle  $2\Theta_0=\cos^{-1}(1/\sqrt{5})$ is defined by
the geodesic arc between two neighboring vertices of the icosahedron.  The
rotations to the nearest neighbors,  ${\bf f}_i$,
from the atom at ${\bf e}$ can be given by the Euler angles
($\alpha=0,\beta=2\Theta_0,\gamma=\pi$), ($\alpha=2\pi/5,
\beta=0,\gamma=0$), and $(\alpha=-2\pi/5,\beta=0,\gamma=0$),
respectively for $i=1,2,3$.

The LUMO's and HOMO's of interest belong to the representations $t_1$ and $h$
respectively, which are isomorphous to the spherical harmonics
$l=1$ ($p$ wave), and $l=2$ ($d$ wave). The state degeneracy is then $2l+1$ and
the normalization $N_l=\sqrt{(2l+1)/60}$.  The corresponding Hamiltonians for
these states are
\begin{eqnarray}
\hat{h}^1 &=& \left(
\begin{array}{c c c}
-1+\frac{2}{\sqrt{5}}  & -\sqrt{\frac{2}{5}} &
-\frac{1}{2}\left(1-\frac{1}{\sqrt{5}}\right) \\
-\sqrt{\frac{2}{5}}    & 2+\frac{1}{\sqrt{5}} & \sqrt{\frac{2}{5}} \\
-\frac{1}{2}\left(1-\frac{1}{\sqrt{5}}\right) & \sqrt{\frac{2}{5}}
& -1+\frac{2}{\sqrt{5}} 
\end{array} \right),  \nonumber \\
\hat{h}^2 &=&  
\left(
\begin{array}{c c c c c}
-\frac{1}{5}-\frac{2}{\sqrt{5}} &
\frac{1}{\sqrt{5}}+\frac{1}{5} &
\frac{\sqrt{6}}{5} &
\frac{1}{\sqrt{5}}-\frac{1}{5} &
\frac{3}{10}-\frac{1}{2\sqrt{5}} \\
\frac{1}{\sqrt{5}}+\frac{1}{5} &
-\frac{1}{5}+\frac{2}{\sqrt{5}} &
-\frac{\sqrt{6}}{5} &
-\frac{3}{10}-\frac{1}{2\sqrt{5}} &
-\frac{1}{\sqrt{5}}+\frac{1}{5} \\
\frac{\sqrt{6}}{5} &
-\frac{\sqrt{6}}{5} &
\frac{9}{5} &
\frac{\sqrt{6}}{5} &
\frac{\sqrt{6}}{5} \\
\frac{1}{\sqrt{5}}-\frac{1}{5} &
-\frac{3}{10}-\frac{1}{2\sqrt{5}} &
\frac{\sqrt{6}}{5} &
-\frac{1}{5}+\frac{2}{\sqrt{5}} &
-\frac{1}{\sqrt{5}}-\frac{1}{5} \\
\frac{3}{10}-\frac{1}{2\sqrt{5}} &
-\frac{1}{\sqrt{5}}+\frac{1}{5} &
\frac{\sqrt{6}}{5} &
-\frac{1}{\sqrt{5}}-\frac{1}{5} &
-\frac{1}{5}-\frac{2}{\sqrt{5}}
\end{array}
\right)  \nonumber \\
\label{ham}
\end{eqnarray}
where $\cos(2\Theta_0)=1/\sqrt{5}$ and
$\cos(\frac{2\pi}{5})=\frac{1}{4}(-1+\sqrt{5})$ are used.

The eigenvalues $\lambda$ of the reduced Hamiltonians for the two LUMO and two
HOMO levels are, for $l=1$,
\begin{eqnarray}
\lambda_{1+}&=&\frac{1}{2}(-3+\sqrt{5}), \nonumber \\
\lambda_{1-}&=&\frac{1}{2}[(3+\sqrt{5})/2-\sqrt{(19-\sqrt{5})/2}] , 
\label{eq2.5}
\end{eqnarray}
and for $l=2$,
\begin{eqnarray}
\lambda_{2-}&=&\frac{1}{2}(-1+\sqrt{5}), \nonumber \\
\lambda_{2+}&=& 1,  \label{eq2.6}
\end{eqnarray}
 in agreement with other calculations \cite{RF-92,DAB-73,YD-92}.
These orbitals are associated, respectively, with the symmetry $t_{1g}$,
$t_{1u}$, $h_u$, and $h_g$. The single-particle energies of these molecule
states are
\begin{equation}
\varepsilon_{lp}=-\lambda_{lp}V, \label{energy}
\end{equation}
where $lp$ runs over the indices 1$+$ (for the representation $t_{1g}$), 1$-$
($t_{1u}$), 2$-$ ($h_u$), and 2$+$ ($h_g$).

For these parity doublets with not so very different energies, the Hamiltonians
(\ref{ham}) give the normalized eigenvectors with $(2l+1)$ components (with
$x=\lambda_{1-}-2-\frac{1}{\sqrt{5}}$ and $\tan y=\sqrt{5}x/2$) at carbon site
${\bf e}$ as
\begin{eqnarray} 
\hat{e}_{1+}({\bf e})&=&\frac{1}{\sqrt{2}}\left(
\begin{array}{c}
1 \\ 0 \\ 1
\end{array}
\right),\hspace*{0.3cm}
\hat{e}_{1-}({\bf e})=\frac{1}{\sqrt{2}}\left(
\begin{array}{c}
-\sin y \\ \sqrt{2}\cos y \\ \sin y
\end{array}
\right), \nonumber \\
\hat{e}_{2-}({\bf e})&=&\frac{1}{\sqrt{10}}\left(
\begin{array}{c}
1 \\ 2 \\ 0 \\ 2 \\ -1
\end{array}
\right),\hspace*{0.3cm}
\hat{e}_{2+}({\bf e})=\frac{1}{\sqrt{30}}\left(
\begin{array}{c}
-1+\sqrt{5} \\ 1+\sqrt{5} \\ \sqrt{6} \\ -1-\sqrt{5} \\ -1+\sqrt{5}
\end{array}
\right), \nonumber\\
   \label{wf}
\end{eqnarray}
triplets for the LUMO states and quintets for the HOMO states.
The complete eigenvectors with the components for the other atoms 
follow by rotation (2.2).
Using this equation, the definition of the Wigner-$D$ functions \cite{DAV-88},
and the form of the vectors in Eq. (\ref{wf}) one can easily show the parity of
the states, $e_{lm\pm}(-{\bf g})=\pm e_{lm\pm}({\bf g})$. The pairs of vectors
for the atoms at the sites ${\bf e}$ and $-{\bf e}$ give instructive examples.
The corresponding transformation matrices $D^l_{mm'}(0,\pi,0)=(-1)^{l+m}
\delta_{m,-m'}$ gives eigenvectors at $-{\it{\bf e}}$, which fulfill the parity
condition. Since the rigid-body transformation from ${\bf e}$ to $-{\bf g}$
may be related to a product of transformations from ${\bf e}$ to ${\bf g}$
and ${\bf e}$ to $-{\bf e}$, the above property is also valid for arbitrary
atomic positions ${\bf g}$.

Fig.~2 shows schematically the two LUMO levels and two HOMO levels and their
associated states. The energy scale is set by the hopping matrix element $V$ in
Eq.~(\ref{energy}).  The single-particle energy difference
$(\varepsilon_{1-}-\varepsilon_{2-})$ is taken to be 3.5 eV between $t_{1u}$
and $h_u$ peaks in solid $C_{60}$ measured by the photoemission and
inverse-photoemission experiments \cite{RWL-92,Taka-92,Weaver}. This yields an
estimate of $V=4.626$ eV.  The brackets $V=$~(3.83, 6.61)~eV represent the
uncertainty of this estimate. The lower value arising out of the finite band
widths is taken to be the midpoint between the band onset at 2.3 eV
\cite{RWL-92} and the peak-to-peak difference at 3.5 eV. The higher value of
$V$ comes from the estimate of 5~eV as the difference between the electron
affinity  level and the ionization potential of the $C_{60}$ molecule
\cite{PJ-91}. The three values of $V$
yield the single-particle excitation spectrum
$\varepsilon_{1+}=4.63$ (3.83, 6.61) eV $(t_{1g})$, $\varepsilon_{1-}=3.50$
(2.90, 5.00) eV $(t_{1u})$, $\varepsilon_{2-}=0$ eV $(h_u)$, and
$\varepsilon_{2+}=-1.77$ (-1.46, -2.52) eV
$(h_g)$ with respect to the position of the
highest occupied state.  The discrepancy between these values of $V$ and the
LDA derived $V= 2.72$~eV \cite{NL-95} represents the phenomenological fit of
the former to the renormalized one-particle energies so that, when the
interaction between two single-particle excitations is considered later, the
one-particle energies should not be further modified by the interaction.

\section{Excitons}
\label{sec3}

\subsection{Symmetry-adapted electron-hole pair states}

In this paper, we shall consider only electron-hole excitations without spin flip,
i.e. only singlet excitons.  We have calculated the energies of the triplet
excitons, which, devoid of exchange interaction terms,
lie slightly lower than the singlets \cite{Shirley}.  The possibility of magnetism
involving the triplet state will be left to a future study.

Consideration of the electron-hole pair excitations of the $\pi$-electron
system of $C_{60}$, which are lowest in energy, can be restricted to the
level scheme of Fig.~2 with the empty levels $t_{1g}$ and $t_{1u}$ and the
occupied states $h_u$ and $h_g$. The pair excitations contain products of the
type $t_{1p_{e}}\times h_{p_{h}}$ with the single-particle parities
$p_e, p_h=+1$ $(g)$ or $=-1$ $(u)$. With the pair parity
$P=p_e\cdot p_h$, where $P$ runs over the same values $+1$ $(G)$ and 
$-1$ $(U)$ as $p_e$ and $p_h$, the pair states have the symmetry \cite{ILLU-89}
\begin{equation}
t_{1p_{e}}\times h_{p_{h}}=T_{1P}+T_{2P}+G_P+H_P.
\end{equation}
That is, each of these four 15-dimensional product representations of
a singlet electron-hole pair from LUMO/HOMO of the type $h_{p_{h}}\rightarrow
t_{1p_e}$ splits up into two three-dimensional representations, $T_{1P}$ and
$T_{2P}$, one four-dimensional $G_P$ representation and one  five-dimensional
$H_P$ representation $(P=G,U)$ \cite{IN}. Among them is the
dipole-allowed pair excitations $T_{1U}$ for electron and hole with opposite
parity.  For these optically observable excitons, there is no need to consider the
fourfold degenerated $g_g$ hole level which is either somewhat below the $h_g$
level \cite{ILLU-89} or degenerate with it within the approximations considered
\cite{RF-92,DAB-73,YD-92}, from the relation \cite{ILLU-89}
\begin{equation}
t_{1p_{e}}\times g_g=T_{2P}+G_P+H_P  ,
\end{equation}
which contains no representation $T_{1U}$ of the electric-dipole-allowed
excitons.
Symmetry reasons also dictate that there is no configurational interaction
between the $A_G$ ground state and the  low-lying pair states considered.

We develop a method of computing the excitonic states in terms of the
symmetry-adapted electron-hole pair states (and incidentally gained some
physical insight into these pair states) by exploiting the close relation of the
representations of the symmetry group $I_h$ of the $C_{60}$ molecule with the
transformation properties of the spherical harmonics \cite{YD-92} which differ
by a small perturbation.  The $(t_{1p_{e}})$ of the LUMO and the $(h_{p_{h}})$ of
the HOMO correspond to single-particle angular momentum states with the
quantum numbers $l$  and $m$ ($l=1,2$ and $-l\le m\le l)$.  The angular
momentum addition rules would yield the symmetry of the resulting
electron-hole pair states  to be those of the spherical harmonics $L$ and $M$
with $L=1,2,3$ and $-L\le M\le L$.  Indeed, the pair states have a
three-dimensional representation $T_{1P}$ corresponding to $L=1$ and a
five-dimensional representation $H_P$ corresponding to $L=2$.  However,  since
$C_{60}$ does not have complete  spherical symmetry, the $L=3$ states split
into two groups, a three-dimensional representation $T_{2P}$ and a
four-dimensional one $G_P$.  The symmetry-adapted electron-hole pair
spin-singlet states may be written as linear combinations
\begin{equation}
\big|LNp_ep_h\big>=\sum^1_{m_e=-1}\sum^2_{m_h=-2}C^{12}_{m_{e}m_{h}}(LN)
c^+_{1p_{e}m_{e}}c_{2p_{h}m_{h}}|0>,  \label{cleb}
\end{equation}
where the operator $c^+_{lpm}$ $(c_{lpm})$ creates (annihilates) an 
electron in a molecule state $|lpm>$ with a single-particle parity $p$. 
The coefficients on the right of Eq.~(\ref{cleb}) are related to the 
Clebsch-Gordan coefficients $C^{LM}_{lml'm'}$ \cite{DAV-88} by
($L=1,2$ with $-L\le N\le L$ and $L=3$ with $N=0,\pm 1$)
\begin{eqnarray}
C^{12}_{m_{e}m_{h}}(LN) &=& (-1)^{-m_{h}}C^{LN}_{1m_{e}2-m_{h}},
\nonumber \\
C^{12}_{m_{e}m_{h}}(3,\pm 3) &=& (-1)^{-m_{h}}\left[\sqrt{\frac{2}{5}}
C^{3\pm 3}_{1m_{e}2-m_{h}} \right. \nonumber \\ 
 & &\left.~~~~~~~~~~~~~~~\pm \sqrt{\frac{3}{5}}
C^{3\mp 2}_{1m_{e}2-m_{h}}\right],  \nonumber \\
C^{12}_{m_{e}m_{h}}(3,\pm 2) &=& (-1)^{-m_{h}}\left[\sqrt{\frac{2}{5}}
C^{3\mp 2}_{1m_{e}2-m_{h}} \right. \nonumber \\
 & &\left.~~~~~~~~~~~~~~~\mp
\sqrt{\frac{3}{5}}C^{3\pm 3}_{1m_{e}2-m_{h}}\right]. 
\label{coeff}
\end{eqnarray}
The symmetry-adapted pair states are chosen such that quantum numbers $L$
and $N$, with $L=1$ corresponds to the basis of the irreducible representation
$T_{1P}$, $L=2$ to $H_P$, $L=3$ and $N=0, \pm 3$ to $T_{2P}$ and $L=3$ and
$N= \pm 1, \pm 2$ to $G_P$.

The true singlet exciton states are linear combinations of the
symmetry-adapted pair states (\ref{cleb})
\begin{equation}
\big|LNP\Lambda\big>=\sum_{p=+,-}c_{\Lambda p}(LNP) \;
\big|LNP{\cdot}p \ p\big>, \label{truepair}
\end{equation}
where the summation runs over the hole parity $p$.  The summation in
Eq.~(\ref{truepair}) indicates that pair states of different single-particle
parities may be coupled provided that the total parity $P$ is conserved.
The fourth quantum number $\Lambda$ labels the two coupled pair states
of the same symmetry. The four quantum numbers, $L$, $N$, $P$, and $\Lambda$
span the 60 pair states (without spin) under consideration.
The eigenstates (\ref{truepair}) of the Frenkel excitons are orthonormalized with
$\sum_pc^*_{\Lambda p}c_{\Lambda 'p}=\delta_{\Lambda\Lambda '}$, following
the orthonormalization property of the  Clebsch-Gordan coefficients.

The symmetry-adapted basis pair states $|LNP{\cdot}p~p>$ of Eq.~(\ref{cleb})
with the third quantum number set to $p_e=P{\cdot}p$ block-diagonalize the
two-body Hamiltonian of the $\pi$-electron system including the full
Coulomb interaction $v$ into $2\times 2$ matrices diagonal in the
quantum numbers $L$, $N$, and $P$: 
\begin{eqnarray}
& & \left<LNP{\cdot}p \ p\big|H\big|L'N'P'{\cdot}p' \ p'\right> \nonumber \\
&=&\delta_{LL'}\delta_{NN'}\delta_{PP'}\left\{\delta_{pp'}
\left[\varepsilon_{1P\cdot p}-\varepsilon_{2p}\right]\right.
\nonumber \\
& &\left.  +\left<LNP{\cdot}p \ p\big|v\big|LNP{\cdot}p' \ p'\right>\right\},
\label{secular}
\end{eqnarray}
with the electron in the level with the excitation energy $\varepsilon_{1p_{e}}$
and the hole in the level with $\varepsilon_{2p_{h}}$ and with the Coulomb
interaction connecting pair states with $(p_e, p_h)$ and $(-p_e, -p_h)$. 
 In our notation system, the exciton energy eigenvalues $E_{LNP\Lambda}$
are independent of $N$ for $L=1,2$, i.e. $(2L+1)$-fold degenerate.
For $L=3$, the exciton energies for the two symmetry sets of $N=0,\pm 3$ and
of $N=\pm 1,\pm 2$ are different but degenerate within each set.

\subsection{Coulomb interaction}

The Coulomb interaction term in Eq.~(\ref{secular}) includes the electron-hole
attraction, and the exchange term to the electron-hole attraction
\cite{ShamRice} with the diagrammatic representation in Fig.~3. 
The exchange terms of Fig.~3b only exist for the spin-singlet exciton.
Assuming non-overlap of the $p_z$-orbitals from different carbon sites of the
$\pi$-like molecule states, we express the Coulomb term in Eq.~(\ref{secular})
in terms of the single-particle eigenstates of Eqs.~(\ref{2.2}) and (\ref{wf}) as
\begin{eqnarray}
& &\left<LNp_ep_h\big|v\big|LNp_e'p_h'\right> = \nonumber \\
& &-\frac{15}{(60)^2}
\sum_{g,g'}G^*_{LN}(12p_ep_h|{\bf gg'})v({\bf g}-{\bf g}')
G_{LN}(12p_e'p_h'|{\bf gg'})  \nonumber\\
& &+2\frac{15}{(60)^2}\sum_{g,g'}G^*_{LN}(12p_ep_h|{\bf gg})
v({\bf g}-{\bf g}') G_{LN}(12p_e'p_h'|{\bf g'g'}) \nonumber \\
 \label{eq3.7}
\end{eqnarray}
with 
\begin{eqnarray}
\lefteqn{G_{LN}(l_el_hp_ep_h|{\bf gg'})=} \nonumber \\
& &\sum^{+l_{e}}_{m_{e}=-l_{e}}
\sum^{+l_{h}}_{m_{h}=-l_{h}} C^{l_{e}l_{h}}_{m_{e}m_{h}}(LN)
e^*_{l_{e}p_{e}m_{e}}({\bf g})e_{l_{h}p_{h}m_{h}}({\bf g'}),
\label{eq3.8}
\end{eqnarray}
where the coefficients $C^{l_{e}l_{h}}_{m_{e}m_{h}}(LN)$ are defined in 
Eq. (\ref{coeff}). The Coulomb potential takes the form
\begin{equation}
v({\bf g})=U\delta_{g0}+\frac{e^2}{\epsilon R_0}\frac{1}{|{\bf g}|}(1-
\delta_{g0}),  \label{coulomb}
\end{equation}
where $U$ denotes the one-site Coulomb matrix element of the $p_z$ orbitals.
$R_0$ is the distance from a carbon atom to the center of the cage and
$\epsilon$ denotes a dielectric constant representing the screening of
the interatomic (but intramolecule) Coulomb interaction. 

The first term on the right of Eq.~(\ref{eq3.7}) comes from the direct
electron-hole attraction and the second is the exchange counter part.  The
factor of 2 may be viewed as originating from the spin degeneracy or the
structure of the singlet.  The double sums over the carbon sites in expression
(\ref{eq3.7}) may be reduced to single sums, by means of the product relation for
the Wigner $D$-functions representing the two rigid-body transformations to the
sites.   Thus, with a simplifying definition for the Coulomb matrix element:
\begin{eqnarray}
& &V_{pp'}(LNP)\equiv -\left<LNP{\cdot}p \ p|v|LNP{\cdot}p' \ p'\right> 
\label{eq3.10} \\ 
&=&-\frac{U}{60}F_{pp'}(LNP)+\frac{e^2}{\epsilon R_0}\left[H_{pp'}(LNP)
-2X_{pp'}(LNP)\right], \nonumber
\end{eqnarray}
which the $N-$dependence serves only to differentiate between $T_{2P}$ and
$G_P$ in the $L=3$ case. For $L=1,2$ ($-L\le N\le L$) and $L=3$ ($N=0,\pm 1$),
 the intraatomic Coulomb interaction is given by
\begin{eqnarray}
\lefteqn{F_{pp'}(LNP)=} \label{intra} \\
& &\frac{15}{2L+1}\sum^{+L}_{M=-L}G^*_{LM}(12P{\cdot}p
\ p|{\bf ee})G_{LM}(12P{\cdot}p' \ p'|{\bf ee}) \nonumber
\end{eqnarray}
and the interatomic Hartree and exchange contributions
\begin{eqnarray}
H_{pp'}(LNP)& &=\frac{1}{4(2L+1)}\sum^{+L}_{M=-L}{\sum_g}'   \label{hartree}
\\ & &G^*_{LM}(12P{\cdot}p\ p|{\bf ge})\frac{1}{|{\bf g}-{\bf e}|}
G_{LM}(12P{\cdot}p' \ p'|{\bf ge}), \nonumber\\
X_{pp'}(LNP)& &= \frac{1}{4(2L+1)}\sum^{+L}_{M=-L}{\sum_g}' \label{exchange}
\\ & &G^*_{LM}(12P{\cdot}p \ p|{\bf gg})\frac{1}{|{\bf g}-{\bf e}|}
G_{LM}(12P{\cdot}p' \ p'|{\bf ee}),  \nonumber
\end{eqnarray}
where the functions $G_{LM}$ are defined in Eq.~(\ref{eq3.8}). The intraatomic
term, Eq.~(\ref{intra}), includes both the Hartree and exchange contributions.

In Table I are listed the values of these three terms evaluated from
Eqs.~(\ref{intra}--\ref{exchange}). The corresponding results using the continuum
approximation of Ref.~\onlinecite{RF-92} differ  little for the Hartree 
contributions, but up to 50 \% for the exchange terms. Table I indicates that the
interatomic electron-hole exchange may be neglected in comparison with the
interatomic electron-hole attraction. This is in complete contrast to the
intraatomic case, which is exchange-dominated since the singlet exciton has
twice the number of exchange terms of equal magnitude as the direct attraction
(cf.\  Fig.~3).  Since the prefactor $U/60$ is smaller than $e^2/(\epsilon R_0)$, it
is evident that the diagonal elements, $V_{pp}(LNP)$, in Eq.~(\ref{eq3.10}) are
dominated by the interatomic Hartree matrix elements, and are, therefore,
positive. The off-diagonal elements $p\not= p'$ are strongly influenced by the
intraatomic exchange. They are often negative (except for $T_{1G}$ and $H_U$).
Table I also shows that a contact-potential approximation, where the
interatomic Coulomb interactions are neglected, is invalid.

\subsection{Pair excitation energies and ground-state stability}

With the Coulomb interaction given in Table I, the 2$\times$2 eigenvalue
problems for the 60 Frenkel excitons originated from the closest
$\pi$-electron-related HOMO and LUMO single-particle states can be solved.
For a given representation $T_1$, $H$, $T_2$, or $G$ and a given total parity
$P=\pm 1$, the 2$\times$2 Hamiltonian for hole states of parities
$p$ and $p'$ may be written as
\begin{eqnarray}
H_{pp'}=(\bar{E}+p\Delta)\delta_{pp'}-\tilde{U}(1-\delta_{pp'}),
\end{eqnarray}
where
\begin{eqnarray}
\bar{E}&=&\frac{1}{2}(E_++E_-), \nonumber \\
\Delta&=&\frac{1}{2}(E_+-E_-), \nonumber \\
\tilde{U} &=& V_{+-}(LNP),  \nonumber \\
E_p &=&\varepsilon_{1P\cdot p}-\varepsilon_{2p}-V_{pp}(LNP). \label{eq3.14}
\end{eqnarray}

The two exciton eigenvalues $E_{LNP\Lambda}$ with $\Lambda=\pm 1$ and the 
corresponding eigenvectors follow as
\begin{eqnarray}
E_{LNP\Lambda}&=&\bar{E}+\Lambda\sqrt{\Delta^2+\tilde{U}^2}, \nonumber \\
c_{\Lambda P}(LNP)&=&\delta_{\Lambda\cdot p,+}\,\cos\eta
+\delta_{\Lambda\cdot p,-}\,  p\cdot \sin\eta , \label{cmix}
\end{eqnarray}
where
\begin{eqnarray}
\sin(2\eta)=\tilde{U}/\sqrt{\Delta^2+\tilde{U}^2}.
\end{eqnarray}
The coupling of the Frenkel excitons with the same representation and parity $P$
 destroys the simple picture that
the reduction of the difference of the single-particle
energies $\varepsilon_{1P\cdot p}-\varepsilon_{2p}$ in Eq.~(\ref{eq3.14})  by
$V_{pp}$ defines the binding energy of the  exciton. The ratio
$\tilde{U}/\sqrt{\Delta^2+ \tilde{U}^2}$ determines the strength of the
redistribution of the two  coupled excitons with $\Lambda=\pm 1$. 
Therefore, the sign of $\tilde{U}$ plays an important role for the actual
oscillator strength for the excitations of electron-hole pairs with different
$\Lambda$ as will be discussed in Sec.~\ref{lin-abs}.

In Figs.~4 and 5 a selected set of pair excitation energies are plotted versus the
strength of the interatomic Coulomb interaction $e^2/(\epsilon R_0)$ for
two different values of the intrasite Coulomb matrix element $U$.
To avoid clutter in Fig.~4, only plotted are excitons of
representations $T_{1P}$ ($L=1$) and $H_P$ ($L=2$), i.e.,
a total of eight exciton energies with $P=\pm 1$ and $\Lambda=\pm 1$. The
other eight exciton energies for the representations $T_{2P}$ and $G_P$
($L=3$) are not plotted  since they do not usually  appear in the optical
spectra considered below. Figure~5 compares the four lowest pair excitations
$T_{1G}$, $T_{2G}$, $G_G$, and $H_G$ with the $A_G$ Hartree-Fock ground state
with even parity. The two intraatomic interaction values chosen are $U=0$ and
$U=4V$.  The Coulomb energy $U$ when two
electrons are in the same atomic $p$-orbital is usually estimated to be
$U\approx$ 10--20~eV. If one uses a hydrogenlike wave function with
an effective nuclear charge $z_{eff}=3.25$, a value of
$U=17.3$ eV results \cite{RF-92}.
With reasonable values for the hopping parameter $V$ (cf. Sec.~\ref{sec2}),
$U=4V$ is close to the estimated values. Both Figs.~4 and 5 show that the
explicitly chosen $U$ value has a minor influence on the excitation energies. A
strong intraatomic Coulomb interaction $U=4V$ shifts the pair energies slightly
to higher energies by about $0.05 V$.

The dependence on the variation of the interatomic Coulomb interaction 
$e^2/(\epsilon R_0)$ is explored because of the uncertainty of the value for the
dielectric constant used to screen the interatomic interaction.  The range goes
 from the limit of metallic screening ($\epsilon=\infty$) to the limit of the
unscreened Coulomb interaction, ($\epsilon=1$). With
$R_0\approx 3.5 \AA$ and the range of the hopping parameter $V$ in
Sec.~\ref{sec2}, the maximum value of  $e^2/(\epsilon R_0V)$ is about 1.1. In
the literature,  screening values between 3 and 10 have been reported.
For instance, dielectric constants for solid fullerites have been determined as
$\epsilon=3.5$, 3.9, or 4.4 \cite{SI-92,PE-92,YNX-91}. A dielectric constant in a
model cluster of 7.13 to 9.86 has been used in Ref.~\onlinecite{PL-92} to study
the van-der-Waals cohesion energy. Other authors \cite{GC-95} use
$\epsilon=4.4$ and 6.5 to explain the screening in $C_{60}$ clusters. Hansen et
al. \cite{PLH-91} reported a value $\epsilon=4.6$ derived from a Kramers-Kronig
analysis of their visible-UV EELS spectrum.  For $\epsilon=4.6$, the interaction
parameters of $e^2/(\epsilon R_0V)\approx 0.19$ (0.23, 0.14) result from
$V=4.626$ (3.83, 6.61) eV. From Fig.~4, evidently the interatomic Coulomb
interaction has
 much more influence on the pair-excitation energies than the onsite Coulomb
interaction. Moreover, its presence gives rise to an effective attractive
interaction between electrons and holes. The exciton binding noticeably reduces
the pair excitation energies. Typical reductions amount to about 
$0.8 e^2/\epsilon R_0$ for $T_{1G}$,
$H_G$, $H_U$, and $G_G$ and vanishing intraatomic interaction
$U$. For $T_{1U}$ ($T_{2G}$), slightly smaller (larger) values of 0.5 (1.0)
$e^2/\epsilon R_0$ are shown in Figs.~4 and 5. The reductions corresponding to
finite $U$ values are smaller. 

The drastic reduction of the electron-hole pair excitation energies by the
 interatomic Coulomb interaction is illustrated in Fig.~5 for the lowest
excitations with even parity.  The figure indicates that for sufficiently weak
screening, pair energies may even become negative. For $T_{2G}$ this happens
if $\epsilon<6.5$ eV/$V$ (when $U=0$) or $\epsilon <5.5$ eV/$V$ (when $U=4V$).
For $V=3.83$ this corresponds to  $\epsilon<1.7$ or $\epsilon <1.4$. 
For a very strong (i.e. almost unscreened) interatomic Coulomb interaction,  
the lowest $T_{2G}$ pair excitation replaces the
closed-shell Hartree-Fock state $A_G$ as the ground
state of the $C_{60}$ molecule.  This transition should
be accompanied by a spontaneous symmetry breaking to lift the degeneracy
of the lower $T_{2G}$ states with respect to the quantum number $N$. 

The findings in Figs.~4 and 5 are consistent with the results of
quantum-chemical calculations  \cite{ILLU-89,RDB-91,FN-92}. 
The energies of the $^1T_{1G}$, 
$^1T_{2G}$, and $^1G_G$ multiplets are quasi-degenerate, 
within a range of 0.1 eV, at about 0.5~eV  and the $^1H_G$ multiplet is higher, 
separated by about 0.4 eV.  Figure~5 (b) ($U=4V$)
indicates a similar situation for reasonable parameters of
$e^2/(\epsilon R_0V)\approx 0.2\mbox{--}0.5$ discussed above. 

The comparison with the quantum-chemical results may be used to fix our
model parameters. In the unscreened case $\epsilon=1$ the disagreement is
too large. When the effect of the interatomic Coulomb interaction is
somewhat screened, e.g. in the case of $U=17.3$ eV and an intermediate effective
screening of $\epsilon=4.6$, the agreement is improved. For
the three hopping parameters $V=4.626$ (3.83, 6.61) eV 
we find 2.86 (2.53, 3.90) eV
for $T_{1G}$, 3.06 (2.65, 4.24) eV for $T_{2G}$, and 2.75 (2.46, 3.71) eV
for $G_G$, and 3.16 (2.71, 4.41) eV for $H_G$.
Consequently, we conclude that the effective interatomic Coulomb
interaction has to be effectively screened.  For the hopping
parameter $V=3.83$ eV, $\epsilon=4.6$ is the appropriate value. For
the larger hopping parameter $V=4.626$ eV the screening has to be slightly
increased.  The largest hopping parameter $V=6.61$ eV should be excluded since
the dielectric constant becomes unreasonably large. The resulting picture is also
more or less consistent with quasiparticle calculations for solid $C_{60}$
\cite{ELS-93} which yield the lowest excitation energy of 2.15 eV for the
$h_u\rightarrow t_{1u}$ transition. When the electron-hole
interaction is included \cite{CHMZ-95} this value is  reduced
to 1.57 eV giving an exciton binding energy of 0.58 eV. 
This {\it ab initio} band calculation also gives
peak positions for $T_{2G}$, $T_{1G}$, $G_G$, and $H_G$ that are redshifted
in comparison to the peak positions 1.86, 1.94, 2.03, and 2.30 eV in the
fine structure of the forbidden absorption edge of the fullerite 
\cite{CHMZ-95}. An inconsistency among the first-principles calculations is 
the energy  ordering of the $T_{1G}$ and $T_{2G}$ excitations.  
In Refs. \onlinecite{ILLU-89,RDB-91} the symmetry of the lowest 
excitation is $T_{1G}$ whereas other calculations \cite{FN-92,CHMZ-95} 
indicate the $T_{2G}$ level to be the lowest one.  Figure~5 shows that the answer
depends on the relative strengths of the intra- and interatomic Coulomb
interactions. For $U=0$, $T_{2G}$ represents the lowest excited state. For
$U=4V$, this holds only for $e^2/(\epsilon R_0V)\ge 0.7$. In the more interesting
region of lower interatomic values $T_{1G}$ is favored.  For even
smaller values of interatomic interaction, $G_G$  becomes the ground state.

\section{Optical Spectra}
\label{sec4}

\subsection{Optical transitions}

The coupling of light  to the $\pi$-electron system of the $C_{60}$ molecule is 
governed by the polarization operator
\begin{equation}
\hat{P}_\alpha=\sum_{\nu,\nu'}\left<\nu\left|ex_\alpha\right|\nu'\right>
c^+_\nu c_{\nu'} \, , \label{eq4.1}
\end{equation}
where $x_\alpha$ is the $\alpha$-th Cartesian component of the position
operator. With the restriction to the LUMO and HOMO states  described in
Sec.~\ref{sec2} and  a strong localization of the $p_z$ orbitals as
in the case of the description of the Coulomb interaction in Eq.~(\ref{coulomb}),
the dipole matrix elements take the form 
$(\nu\equiv lpm)$:
\begin{eqnarray}
\lefteqn{\left<lpm\left|ex_\alpha\right|l'p'm'\right>=} \nonumber \\
& &eR_0 \frac{\sqrt{(2l+1)(2l'+1)}}{60}
\sum_ge^*_{lpm}({\bf g}){\rm g}_\alpha e_{l'p'm'}({\bf g}). \label{eq4.2}
\end{eqnarray}
The sum over the carbon positions can be evaluated using the symmetry
properties \cite{DAV-88}:
\begin{eqnarray}
\lefteqn{\left<lpm\left|ex_\alpha\right|l'p'm'\right>=} \nonumber \\
& &\delta_{p,-p'} D(ll'|pp')
\sum^{+1}_{M=-1}(-1)^{m'}C^{1M}_{lml'-m'}A^1_{\alpha M}, \label{dipole}
\end{eqnarray}
where the effective dipole moment is given by
\begin{eqnarray}
D(ll'|pp')&=&eR_0\frac{\sqrt{(2l+1)(2l'+1)}}{3} \nonumber \\
& &\times\left[-\sqrt{2}G_{11}
(ll'pp'|{\bf e e}){\rm e}_x+G_{10}(ll'pp'|{\bf e e}){\rm e}_z\right],
\end{eqnarray}
with $G_{LN}(ll'pp'|{\bf e e})$ defined in Eq.~(\ref{eq3.8}) for the reference atom
at position ${\bf e}$, and the projection of the Cartesian components 
onto the $L=1$ angular momentum states,
\begin{eqnarray}
A^1_{\alpha M}&=&\delta_{\alpha x}\frac{1}{\sqrt{2}}\left(\delta_{M-1}-
\delta_{M1}\right) \nonumber \\
&+& \delta_{\alpha y}\frac{i}{\sqrt{2}}\left(\delta_{M1}+
\delta_{M-1}\right)+\delta_{\alpha z}\delta_{M0}.
\end{eqnarray}

The optical transition matrix elements from the $A_G$ ground state to
the electron-hole pair excited states introduced in Eq.~(\ref{cleb}) are
\begin{equation}
\left<LNp_ep_h\left|\hat{P}_\alpha\right|0\right>=\delta_{L1}
\delta_{p_{e},-p_{h}}D(12|p_ep_h)A^1_{\alpha N} \, . \label{eq4.6}
\end{equation}
Selection rules limit excitations to $L=1$ excitons with odd parity, i.e. $T_{1U}$
states \cite{ILLU-89,FN-92}. Out of the 60 singlet excitons there are only six 
such states, namely $|1N+->$ and $|1N-+>$ with $N=-1,0,1$.  Because of the 
different structures of the single-particle eigenvectors in Eq.~(\ref{wf}) the
oscillator strengths vary with the parity-allowed pairs:
\begin{eqnarray}
D(12|+-) & = & -0.48547 eR_0 \nonumber \\
D(12|-+) & = & -0.55626 eR_0.  \label{dipolelength}
\end{eqnarray}
The characteristic dipole length is about half the distance from a carbon atom to
the  center of the cage.

The coupling of two different excitonic states by an external electric field
via the  polarization operator is
\begin{eqnarray}
\lefteqn{\left<LNp_ep_h\left|\hat{P}_\alpha\right|L'N'p_e'p_h'\right>} 
\nonumber \\
 & = & \sum^{+1}_{m_{e},m_{e}'=-1}\sum^{+2}_{m_{h},m_{h}'=-2}
       (-1)^{-m_{h}-m_{h'}}
       C^{LN}_{1m_{e}2-m_{h}} C^{L'N'}_{1m_{e}'2-m_{h}'}  \nonumber \\
 & &\times  \{-\delta_{p_{e}p_{e}'}\delta_{m_{e}m_{e}'}\left<2p_h'm_h'
\left|ex_\alpha\right|2p_hm_h\right> \nonumber \\
& & +\delta_{p_{h}p_{h}'}\delta_{m_{h}m_{h'}}
\left<1p_em_e\left|ex_\alpha\right|1p_e'm_e'\right> \}.
\label{eq4.7}
\end{eqnarray}
We record here a special case needed later for the nonlinear optical spectra, by
Eq.~(\ref{dipole}),
\begin{eqnarray}
& &\left<2Npp\left|\hat{P}_\alpha\right|1N'{-p'} \ p'\right>  \nonumber \\
& = &\sqrt{\frac{3}{20}}\left[\delta_{pp'}\sqrt{5}D(11|p \ {-p})+
\delta_{p,{-p'}}D(22|{-p} \ p)\right]  \nonumber \\
& & \times (-1)^{N'}\sum^{+1}_{M^{''}=-1}C^{1M}_{2N1-N'}A^1_{\alpha M},
\label{eq4.8}
\end{eqnarray}
with $D(11|-+)=D(11|+-)=-0.70700 eR_0$ and $D(22|+-)=D(22|-+)=0.75585 eR_0$. 
The dipole-allowed transitions from $T_{1U}$ to both
$H_G$ and $T_{1G}$ excitons are possible because of the difference of the
$C_{60}$ symmetry group from the spherical symmetry.

\subsection{Linear absorption: Optically allowed excitons}
\label{lin-abs}

Consider first the case of the $A_G$ ground state.  Its
 optical properties are governed by the time-dependent  polarization field
\begin{equation}
P_\alpha(t)=2n\left<0\left|\hat{P}_\alpha(t)\right|0\right>, \label{polar}
\end{equation}
where $n$ is the density of the buckyballs.  The linear response in the
rotating-wave approximation  is, following Eq.~(\ref{eq4.6}), given by
\begin{equation}
\chi^{(1)}_{\alpha\beta}(\omega)=\delta_{\alpha\beta}2n\sum_{\Lambda=+,-}
\frac{\left|\sum_{p=+,-}c_{\Lambda p}(1-)D(12|{-p} \ p)\right|^2}{E_{1-\Lambda}
-\hbar\omega-i\Gamma_{1-\Lambda}}, \label{eq4.11}
\end{equation}
where $\omega$ is the frequency of light and a phenomenological
lifetime-broadening parameter $\Gamma_{LP\Lambda}$ has been introduced for
the electron-hole pair states. 
The diagonality and isotropy of the susceptibility tensor follows immediately 
from $\sum^{+1}_{M=-1} A^{1^{*}}_{\alpha M} A^1_{\beta M} =
\delta_{\alpha\beta}$. 

The resulting low-energy absorption spectrum of the $\pi$-electron system is
shown in Fig.~6 for various values of the effective  intraatomic Coulomb
interaction $U$. The ratio of the interatomic and intraatomic Coulomb
interaction has been fixed at the values 
$e^2/(\epsilon R_0U)=0.2377$ (left panel) and 0.0517 (right panel). Using
an intrasite matrix element $U=17.3$ eV the two values correspond to the
cases of no screening of the interatomic interaction ($\epsilon=1$)
and of intermediate screening ($\epsilon=4.6$).
In the spectral range considered the absorption
in the $\pi$-electron system exhibits two $T_{1U}$ exciton peaks at about 
$E_{1- -}\approx V+0.012U-0.666\frac{e^2}{\epsilon R_0}$ 
and $E_{1-+}\approx 1.139 V$ 
$+0.039U-0.488\frac{e^2}{\epsilon R_0}$ with different oscillator strengths.

Comparison of the calculated absorption spectrum with experiment leads to
a discussion of two salient points: (i) the comparison of the calculated and 
measured peak positions and (ii) the relative intensities of the two absorption
peaks.   The two calculated exciton peaks should be compared to  the measured
absorption peaks at 3.81 and 4.90 eV \cite{ZG-91}, or 3.65 and 4.72 eV
\cite{SLR-91}, or 3.78 and 4.84 eV \cite{HA-90}. The weak structure observed at
lower energies,  in particular the weak peak at 2.73 eV, for solid $C_{60}$
\cite{SLR-91} has been ascribed to dipole-forbidden transitions, which become
partially allowed due to lattice fluctuations, interface effects and/or internal
electric fields. The same holds for the small structure occurring in the energy
region of 2 eV in spectra of
$C_{60}$ isolated in a noble gas matrix \cite{ZG-91}. The intense absorption band
close to 5.96 eV \cite{SLR-91} or 5.87 eV \cite{HA-90} only possesses a  partial
$\pi$ character and is also beyond the scope of our study.  The identification of
the lower and upper $T_{1U}$ excitons ($L=1$, $P=-$) with the two absorption
peaks under consideration restricts the range for the values of the parameters
$U$ and $e^2/(\epsilon R_0)$. If we take the strong onsite interaction value
$U=17.3$ eV, the dielectric constant varies drastically between
$\epsilon\approx 2$ (for the hopping parameter $V=4.62$ eV) and
$\epsilon\approx 10$ (for $V=3.83$ eV).  With a weaker $U\approx 10$ eV, the
range of the dielectric  constant is reduced to $2.5<\epsilon <4.8$.

The theoretical result that the absorption peak at lower energy possesses a
smaller oscillator strength than the  peak at higher energy is in
agreement with the experimental observation.
Within the single-particle approximation, i.e., for $U= 0$ and 
$e^2/(\epsilon R_0)= 0$, the oscillator  strengths of the two transitions
$h_u\rightarrow t_{1g}$ and $h_g\rightarrow t_{1u}$ between molecule levels
are similar, from Eq.~(\ref{dipolelength}). 
The reduction of the ratio of the oscillator strength of the lower-energy peak to
that of the higher-energy peak may be understood in terms of the Coulomb
coupling of the two $T_{1U}$ excitons, as is evident from the presence of the 
eigenvectors $c_{\Lambda p}(1-)$ from Eq.~(\ref{cmix}) in the oscillator
strength of Eq.~(\ref{eq4.11}).  Since the dipole matrix elements $D(12|+-)$ and
$D(12|-+)$ have the same sign and nearly equal magnitudes,   
 the  oscillator strengths of the two $T_{1U}$ excitations are governed by
the relative sign of the coefficients $c_{\Lambda p}(1-)$, and so by the sign of 
the coupling term $V_{+-}(1-)$ in Eq.~(\ref{eq3.10}).  The domination of the
exchange term leads to $V_{+-}(1-)<0$ and, hence, the oscillator strength of the
low-energy exciton at $\hbar\omega=E_{1- -}$ is reduced compared to the
high-energy absorption at $\hbar\omega=E_{1-+}$. 
This shows not only that the relative strengths of the two exciton peaks are
influenced by the  Coulomb interaction but also that the intersite
interaction is indispensable. 

Let us consider briefly the consequences on the linear optical spectra of the
fascinating possibility raised in Sec.~\ref{sec3} that the excitonic state
$T_{2G}$ is the ground state rather than the  closed-shell $A_G$ state. In the
 energy region of the two allowed single-particle transitions
$h_u\rightarrow t_{1g}$ and $h_g\rightarrow t_{1u}$,  two peaks can arise
from the transitions between the lowest $T_{2G}$ state to the two $H_U$
and two $G_U$ states (see Fig.~4) satisfying both the angular momentum and
parity selection rules. Their oscillator strengths are governed by the moments
$D(11|+-)$ and $D(22|-+)$. The transition energies may also be
accounted for by the parameter values for $V$, $U$, and $e^2/(\epsilon R_0)$ 
already discussed in connection with Fig~5.  
However, the single-particle energies from the new excitonic ground state have
to be  recalculated and compared with the measurements by photoemission and
inverse photoemission.  Since in the exciton ground state there is already one
electron in a LUMO and a hole in a HOMO state, the one-particle excited state has
strong correlation effects.  The interaction effects can split the degeneracies of
the HOMOs and the LUMOs, leading to more than two electron and two hole
single-particle states.  The group-theoretical indentities $t_{1p}\times
T_{2G}=g_p+h_p$ and $h_p\times T_{2G}=t_{1p}+t_{2p}+g_p+h_p$ ($p=g,u$) give
us a rough idea of the single-particle multiplets.  The symmetry breaking which
will remove the degeneracy of the $T_{2G}$ states would reduce the number of
one-particle excited states.

\subsection{Electro-optic Kerr effect: Electric-field-induced
forbidden excitons}

In the theoretical linear optical spectra, the parity selection rule excludes the
same-parity transitions, such as those lowest in energy, $h_u\rightarrow
t_{1u}$.  They can, however, be induced by the application of a static electric
field which mixes the  even-parity excitons with the odd-parity
excitons.  The linear response to the external laser field of the
electric polarization,  Eq.~(\ref{polar}), to second order in an applied static
electric field {\bf F} (the Kerr effect), is closely related to the third-order
susceptibility, to be derived in Sec.~\ref{harmonic3}.  By a similar derivation
leading to Eq.~(\ref{chi3}), the static electric field effect on the linear optical
response is given by
\begin{eqnarray}
\lefteqn{\chi^{(1)}_{\alpha\beta}(\omega)=2nP_{\alpha\beta}(\hat{F})
 \frac{F^2}{4}\sum_{L=1}^2 \sum_{\Lambda ',\Lambda ''=\pm}}  \label{eq4.13} \\
& & \frac{S^*_{L\Lambda '}S_{L\Lambda ''}}{(E_{1-\Lambda '}-E_{L+-})
(E_{L+-}-\hbar\omega-i\Gamma_{L+-})(E_{1-\Lambda ''}-E_{L+-})},
\nonumber
\end{eqnarray}
with the prefactor depending on the direction of the applied static field denoted
by its unit vector $\hat{F}$ relative to the light polarization direction:
\begin{equation}
P_{\alpha\beta}(\hat{F}) =\frac{3}{10} (3\delta_{\alpha\beta}+\hat{F}_\alpha
\hat{F}_\beta), \label{prefactor}
\end{equation}
and the oscillator strength $S_{L\Lambda}$ given in Eq.~(\ref{osc3}).

The selection rules $\Delta L=0, \pm1$ confine the contributions to the Kerr
effect from the $H_G$ and $T_{1G}$ excitons.  Because of the random
orientations of the $C_{60}$ molecules in solutions or in the face-centered
cubic thin films \cite{PAH-91,RS-91,WIF-91}, the dependence of the field
direction of the prefactor is in practice averaged out to
$P_{\alpha\beta}(\hat{F})=\delta_{\alpha\beta}$.  
Only in the case of the low-temperature simple cubic crystals of undoped
$C_{60}$ is there a chance of experimentally partially probing the polarization 
dependence.  The four molecules per unit cell are rotated by an angle
$\phi$ around a space diagonal axis [111], [1$\bar{1}\bar{1}$],
[$\bar{1}1\bar{1}$], or [$\bar{1}\bar{1}$1].  The angle of rotation
is  found to be $\phi=22-26^\circ$ \cite{PAH-91,RS-91,WIF-91}. 

The spectrum of the electrooptic Kerr effect is presented in Fig.~7 as a function
of the reduced photon energy for different model parameters of the effective
Coulomb interaction, $U$ and $e^2/\epsilon R_0$. The electric field strength
has been fixed at $eFR_0/(2V)=0.001$ to compare with the allowed 
transitions in zero field in Fig.~6, corresponding to $2.6{\times}10^5$ V/cm for a
hopping parameter of $V=4.626$ eV. 
The positions of the field-induced excitons are approximately given by
$E_{2+-}=0.757V+0.019U-0.760e^2/\epsilon R_0$ for the $H_G$ exciton or
$E_{1+-}=0.757V + 0.0121U - 0.986e^2/\epsilon R_0$ for the $T_{1G}$ exciton.
Their energy difference is determined by the interatomic Coulomb interaction,
$E_{2+-}-E_{1+-}\approx 0.226 e^2/\epsilon R_0 \approx
0.93~\mbox{eV}/\epsilon$.  For intermediate screening, it is smaller than the
line broadening used in Fig.~7, explaining the appearance of only one pronounced
field-induced peak in the absorption spectrum below the energy of the lower
allowed $T_{1U}$ exciton (in Fig.~4).  Only for very weak screening does a weak
second $T_{1G}$-related peak appear at an energy below the more intense
$H_G$-related peak (see the solid and dashed curves in Fig.~7a).

These general findings seem to be in agreement with the results inferred from
different nonlinear optical experiments, such as  two-photon absorption
\cite{GPB-97} and degenerate four wave mixing \cite{FPS-96}. A two-photon
resonance at 2.73 eV or 2.67 eV is observed and identified with an $H_G$
exciton.  Such energy values are within the range of the theoretical values given
by the parameters under discussion.  The prediction of a field-induced $H_G$
exciton line also explains the weak absorption structure around 2.73 eV
\cite{SLR-91}. The theoretical energy ordering $T_{1G}<H_G<T_{1U}$ clearly
visible in Figs. 4 and 5 agrees with results of the second-harmonic generation
(SHG) on a surface \cite{BK-93,AMJ-95}, where a resonance near 1.8--1.9~eV is
identified with a $T_{1G}$ exciton.  However, the theoretical prediction of the
energy splitting between $T_{1G}$ and $H_G$ is smaller than the experimental
value of 0.8--0.9~eV if we choose our model parameters, such as $U=17.3$ eV,
$\epsilon=4.6$, to obtain similar exciton energies as the quantum-chemical
calculations.

\subsection{Third-harmonic generation}
\label{harmonic3}

In this section, we consider another nonlinear optical experiment, the
third-harmonic generation (THG).  A three-photon resonance occurs when
three times the fundamental photon energy is equal to the lowest
$A_G\rightarrow T_{1U}$ one-photon dipole-allowed transition (in the
single-particle picture: $h_u\rightarrow t_{1g}$ and
$h_g\rightarrow t_{1u}$).  From the general expression for the third-order
susceptibility given by Armstrong et al. \cite{JAA-62}, we
take into account only triply resonant contributions to the susceptibility
describing THG. Neglecting biexciton effects \cite{OSS} and using the dipole
matrix elements (\ref{eq4.6}) and (\ref{eq4.8}) we find
\begin{eqnarray}
\chi^{(3)}_{\alpha\beta\gamma\delta}(\omega)=\frac{2n}{3!}
{\rm P}_{\beta\gamma\delta} \sum_{L=1}^2
\sum^{+L}_{N=-L}\sum^{+1}_{N',N''=-1}
\sum_{{\Lambda},{\Lambda}',{\Lambda}'' =\pm 1} \nonumber \\
\left<0\left|\hat{P}_\alpha\right|1N'-\Lambda '\right> 
\frac{\left<1N'-\Lambda '\left|\hat{P}_\beta\right|LN+\Lambda\right>}
{\left(E_{1-\Lambda '}-3\hbar\omega-i\Gamma_{1-\Lambda'}\right)}
\nonumber \\
\frac{\left<LN+\Lambda\left|\hat{P}_\gamma\right|1N''-\Lambda''\right>}
{\left(E_{L+\Lambda}-2\hbar\omega-i\Gamma_{L+\Lambda}\right)}
\frac{\left<1N''-\Lambda''\left|\hat{P}_\delta\right|0\right>}
{\left(E_{1-\Lambda ''}-\hbar\omega-i\Gamma_{1-\Lambda''}\right)}.
\label{eq4.16}
\end{eqnarray}
Here, $n$ is the density [cf. Eq.~(\ref{polar})] and P$_{\beta\gamma\delta}$
denotes the sum over all permutations of
$\beta$, $\gamma$, and $\delta$ ensuring that the fourth-rank tensor
third-order susceptibility is independent of the ordering of those three
indices, i.e., the Cartesian components of the three fields creating the
third-order harmonic.

Selection rules dictate that the three-photon resonance takes the system from
the $A_G$ ground state to the $L=1$ excitons with odd parity ($T_{1U}$), and
from $T_{1U}$ to $L=2$ ($H_G$) and $L=1$ ($T_{1G}$) excitons with even parity,
which may give rise to the doubly resonant terms. We restrict ourselves
to the double-resonance terms with the lower-energy ($\Lambda=-$) 
excitons. The higher-lying  excitons are  energetically well separated
from the frequency region of interest (cf. Fig.~4). 
Consequently the $|LN+-\rangle$ states may be approximately replaced by the
uncoupled $|LNpp\rangle$ ($p=+$, $L=1,2$) states, corresponding to the
single-particle transitions  $h_u\rightarrow t_{1u}$. 
The third-order susceptibility becomes
\begin{eqnarray}
\lefteqn{\chi^{(3)}_{\alpha\beta\gamma\delta}(\omega)
=2nG_{\alpha\beta\gamma\delta}  
\sum_{L=1}^2 \sum_{\Lambda ',\Lambda ''=\pm}}  \nonumber \\
& &\frac{S^*_{L\Lambda '}}{\left(E_{1-\Lambda '}-3\hbar\omega-i
\Gamma_{1-\Lambda '}\right)} \nonumber \\
& &\frac{S_{L\Lambda ''}}{\left(E_{L+-}-2\hbar\omega-i\Gamma_{L+-}\right)
\left(E_{1-\Lambda ''}-\hbar\omega-i\Gamma_{1-\Lambda ''}\right)},
\label{chi3}
\end{eqnarray}
with  the polarization-dependent prefactor
\begin{eqnarray}
G_{\alpha\beta\gamma\delta}&=&\frac{1}{8}{\rm P}_{\beta\gamma\delta}
\sum^{+2}_{M=-2}F^*_{\alpha\beta}(M)F_{\delta\gamma}(M), 
\label{prefactor3} \\
F_{\alpha\beta}(M)&=&\sum^{+1}_{M',M'=-1}(-1)^{M'}A^1_{\alpha M'}
C^{1M''}_{2M1-M'}A^1_{\beta M''},  \nonumber
\end{eqnarray}
and the oscillator strength
\begin{eqnarray}
S_{L\Lambda}&=& \sum_{p=\pm}c^*_{\Lambda p}(1-)D(12|{-p} \ p) 
\nonumber \\
& &\times \left[\frac{1}{\sqrt{5-2L}}c_{\Lambda+}(1-)D(11|+-)\right. 
\nonumber \\
& &+\left. \sqrt{\frac{5-2L}{5}}c_{\Lambda-}(1-)D(22|-+)\right].  
\label{osc3}
\end{eqnarray}
Eq.~(\ref{chi3}) may be interpreted as the third-order
susceptibility of a five-level system: ground state $A_G$, even-parity excited
states $H_G$ or $T_{1G}$, and the two odd-parity  excited states $T_{1U}$, in
contrast to the third-order susceptibility of a three-level system used to fit the
experimental measurements \cite{FK-94,FKC-93}. Moreover, the symmetry of the
states participating in the THG process are well defined here. On the other hand,
only the most dominant resonant terms are considered here, whereas the 
formulas used to fit the data also include less important resonant and
nonresonant terms.

The spectral behavior of the magnitude of the third-order susceptibility is
plotted in Fig.~8 for the same set of model parameters as in Fig.~7. 
Only the lower $T_{1U}$ excitons appear.  Despite the presence of
many exciton levels, in the spectral region considered only a pronounced 
double-peak structure or one broad peak appears.  In the case of
stronger screening of the intersite Coulomb interaction (the right panel of
Fig.~8),  the high-energy peak corresponds to the two-photon resonance with the 
$L=2$, $P=+$ ($H_G$) exciton.  It is enhanced by the resonance with the
$T_{1G}$ excitons ($L=1$, $P=+$). The three-photon resonance at the frequency
of the fundamental light wave occurs at slightly lower energies defined
by the electric-dipole-allowed $L=1$, $P=-$ ($T_{1U}$) exciton. 
 On the other hand, in the case of the weaker screening (the left panel), for the
curves going from the right to the left in order of increasing onsite Coulomb
interaction $U$,  there is an interchange of the energy order of the two
resonances $\hbar\omega=\frac{1}{3}E_{1--}$ and 
$\hbar\omega=\frac{1}{2}E_{2+-}$, $\frac{1}{2}E_{1+-}$.   The near coincidence 
of the triple and double resonances for the intermediate values of $U$ creates
the appearance of a strong single peak in Fig.~8a.    For larger values of the
onsite Coulomb interaction, the $H_G$ and $T_{1G}$ resonances split again and a
weak $T_{1G}$-related peak occurs at the low-energy tail of the THG structure.

Our calculated spectrum can be used to infer the
symmetry of two-particle elementary excitations observed experimentally. In
the THG experiment \cite{FK-94,FKC-93} two peaks were observed at
1.3~$\mu$m  and 1.06~$\mu$m.  By taking a combination of relatively large
parameters $V$, $U$, and $e^2/(\epsilon R_0)$, we can interpret the lower 
two-photon resonance to yield the measured energy of the $H_G$ exciton at
1.9~eV and the higher triple-photon resonance to yield the energy of the
$T_{1U}$ exciton at 3.5~eV.  A previous interpretation of the low-energy peak in
the THG spectrum as a two-photon resonance with the one-photon forbidden
$T_{1G}$ level \cite{FK-94}, even though in agreement with the  low value of the
resonance energy, could not explain the absence of the $H_G$ exciton in the
spectrum.  We have demonstrated here that, with the help of a careful symmetry
analysis, nonlinear optical spectroscopy  can clarify the complicated electronic
structure of the $C_{60}$ molecule, especially its electron-hole pair excitations.

We note briefly that, in the case of the excitonic ground state $T_{2G}$, the
interpretation of the THG would be different. According to the $\Delta L=\pm 1$
and $\Delta L=0$ selection rules, a three-photon resonance excites the $H_U$ or
$G_U$ exciton at $\hbar\omega=(E_{L--}-E_{3+-})/3$, $(L=2,3)$.
Many transitions may contribute to the two-photon resonance. Among them are
$H_U\rightarrow T_{1G}$, $H_G$, $T_{2G}$, and $G_G$, or $G_U\rightarrow H_G$,
$T_{2G}$, and $G_G$.

\section{Discussions}
\label{sec5}

In this paper, we address the issue of the strong electron correlation in a
$C_{60}$ molecule by an analytical model making maximum use of
symmetry.  The basic component is a simple quasi-particle model based on a
tight-binding scheme of $\pi-$orbitals of the HOMO and LUMO states, with
parameters deduced from photoemission and inverse photoemission
experiments.  The assumption of the closed-shell ground state is tested against
the energies of the electron-hole pair excited states.  The interaction between
electrons and holes includes the Coulomb interaction of the
$\pi-$orbitals on the same carbon atom and the long-range interaction
between different carbon sites with a constant dielectric screening. The onsite
electron-hole interaction is dominated by the exchange term, and is, therefore,
repulsive while the electron-hole interaction on different carbon
sites is attractive.  The balance between these two terms depends, in our model, 
on the dielectric screening.  

Our theory of the linear optical absorption
spectra based on the closed-shell ground state agrees with experiment for a
reasonably strong dielectric screening of the long-range interaction.
We have established the relation between the energies or relative oscillator
strengths of the dipole-active excitons and the Coulomb interaction terms
connecting these excitons of the same symmetry.    This coupling
effect explains the observed strong $T_{1U}$ exciton peaks in the linear optical
absorption spectra. 
For the third-harmonic generation, our theory identifies the doublet excitation
at the fundamental wavelength $\lambda\approx 1.3~\mu$m as the forbidden
$H_G$ exciton and not as the $T_{1G}$ exciton suggested by experiment, whereas
the triple resonance is clearly related to the $T_{1U}$ excitation. This
identification is supported by the argument invoking the approximate angular
momentum selection rule in the nearly centrosymmetric molecule (see
Sec.~\ref{sec4}). The same transition to the $H_G$ pairs is found to play a role if
an external static electric field is applied to the $C_{60}$ molecules. We
predict a pronounced optical Kerr effect with a photon resonance at the $H_G$
energy.

We wish to note an interesting scenario in which the closed-shell $A_G$
state is unstable against an electron-hole pair excitation of symmetry $T_{2G}$
with a reasonably weak dielectric screening.  The consequences of having a
$T_{2G}$ exciton state as the ground state are extraordinary.  The optically 
dipole-allowed states are subsets of the $H_U$. The most interesting possibility
is that a quasi-particle state now consists of an electron (or hole) plus an
electron-hole pair, leading to strong correlation effects.  Symmetry
consideration would lead to more photoemission lines than the closed-shell
ground state scenario.  An experimental test of the electro-optic
effect, which can discriminate the ground state symmetry, is described in
Sec.~IV.C.

A less extreme possibility is the small excitation energy of the $T_{2G}$
exciton.  The quasiparticle dynamics can be affected by the easy Coulomb
excitations of such excitons.  In particular, this could provide a source of
effective quasiparticle interaction and, thus, possibly a source for
superconductivity.

While our model study of the excitons has indicated that the most likely
scenario of the ground state in $C_{60}$ is the closed-shell $A_G$, it points out
the possibility of constructing other molecular solids which lowers the excitonic
states.  The requirement is the strength of the inter-site
electron-hole attraction over the onsite repulsion, which might be achieved by
increasing the number of atom sites in a molecule.  Our model approach may
also be used to study excitons in quantum dots when the effective-mass
approximation fails.  We leave the investigation of the above conjectures to
future work.

\section*{Acknowledgment}

One of the authors (F.B.) gratefully acknowledges the kind hospitality of
the University of California at San Diego. This work is financially
supported by the Volkswagen Foundation and the National Science
Foundation (NSF) (Grant No. INT-9513363, DMR-9421966, and DMR-9721444).

\figure{\figurename 1: The structure of the $C_{60}$ molecule.
Each dot denotes the position of a carbon atom. The connecting
solid lines indicate the bonds. The intact icosahedron with the vertices at 
$N$, $S$, $A_i$, and $A_i'$ ($i=1,\ldots,5$) is indicated by the dashed lines.}

\figure{\figurename 2: Schematic four-level diagram of HOMO and LUMO states
for characterization of the lowest electronic excitations in the $C_{60}$
molecule.}

\figure{\figurename 3: Feynman diagrams for the electron (upward arrows) - hole
(downward arrows) interaction (dashed line):  (a) the direct attraction, (b) the
exchange counterpart.}

\figure{\figurename 4: The excitation energies of Frenkel excitons
belonging to the single-particle pairs $h_u\rightarrow t_{1u}$,
$h_u\rightarrow t_{1g}$, $h_g\rightarrow t_{1u}$, and $h_g\rightarrow t_{1g}$
versus the effective
interatomic Coulomb interaction strength $e^2/(\epsilon R_0)$. 
The intraatomic Coulomb interaction is fixed at two values
 (a) $U=0$ and (b) $U=4V$.
$T_{1P}$: solid line, $H_P$: dashed line. Both parities $P=+,-$ are considered.
The high-energy (low-energy) solutions of the coupled
pairs $\Lambda=+$ $(\Lambda=-)$ are plotted as thick (thin) lines.
All energies are given in units of the hopping parameter $V$.}

\figure{\figurename 5: The lowest even-parity exciton energies versus
the interatomic Coulomb interaction. The zero line is given by the 
closed-shell Hartree-Fock ground state. (a) $U=0$, (b) $U=4V$.}

\figure{\figurename 6: Absorption spectra of $C_{60}$ versus the photon
energy in units of $V$ for different parameters of the effective Coulomb
interaction and damping parameters $\Gamma$. Solid line: $U=3V$,
 dashed line: $U=2V$, dotted line: $U=V$, dash-dotted line: $U=0$. The ratio of the
intersite and intrasite Coulomb interaction is fixed to $e^2/(\epsilon
R_0U)=0.2377$ (left panel) and 0.0517 (right panel).}

\figure{\figurename 7: Absorption spectra of $C_{60}$ molecules in a
static electric field versus the  photon energy near the forbidden transition
$h_u\rightarrow t_{1u}$ for different
parameters of the effective Coulomb interaction. Solid line: $U=3 V$,
dashed line: $U=2 V$, dotted line: $U=V$, dash-dotted line:
$U=0$. (a) $e^2/(\epsilon R_0U)=0.2377$, (b) $e^2/(\epsilon R_0U)=0.0517$. 
The damping parameter is chosen as $\Gamma=0.05  V$.
Each spectrum has to be multiplied with the prefactor
$(eFR_0/2V)^2$ to compare with the strength in Fig.~6.}

\figure{\figurename 8: Spectral variation of the THG susceptibility in the
region of the lower 3-photon $T_{1U}$ and 2-photon $H_G$ resonances for the
same parameters of the effective Coulomb interaction as in Fig.~7.  The damping
parameter is chosen as $\Gamma=0.03 V$ for both allowed and forbidden
excitons.}

\newpage
\widetext
\begin{table}
{\scriptsize
  Table I.
  The intraatomic and interatomic Coulomb interaction
  matrix elements (\ref{intra}, \ref{hartree}, \ref{exchange}). 

(I.R. --- Irreducible representation)}
\begin{tabular}{|c|c c c|c c c|c c c|c c c|}
{\scriptsize I.R.} & \multicolumn{3}{c|}{{\scriptsize Quantum numbers}} &
\multicolumn{3}{c|}{{\scriptsize $F_{pp'}(LNP)$}} &
\multicolumn{3}{c|}{{\scriptsize $H_{pp'}(LNP)$}} &
\multicolumn{3}{c|}{{\scriptsize $X_{pp'}(LNP)$}} \\
     & {\scriptsize $L$} & {\scriptsize $N$} & {\scriptsize $P$} &
{\scriptsize $+ +$} & {\scriptsize $- -$} & {\scriptsize $+ -$} & {\scriptsize
$+ +$} & {\scriptsize $- -$} & {\scriptsize $+ -$} & {\scriptsize $+ +$} &
{\scriptsize $- -$} & {\scriptsize $+ -$}
\\ \hline
{\scriptsize $T_{1G}$} & {\scriptsize 1} & {\scriptsize 0,$\pm$1}        &
{\scriptsize 1}  & {\scriptsize 0.50000}
& {\scriptsize 0.76690} & {\scriptsize 0.61923}  & {\scriptsize 0.87251} &
{\scriptsize 0.89689} & {\scriptsize 0.22713}
& {\scriptsize -0.02111} & {\scriptsize -0.03239} & {\scriptsize -0.02615} \\
{\scriptsize $H_G$}    & {\scriptsize 2} & {\scriptsize 0,$\pm$1,$\pm$2} &
{\scriptsize 1}  & {\scriptsize 1.50000}
& {\scriptsize 1.23310} & {\scriptsize 0.79877}  & {\scriptsize 0.85182} &
{\scriptsize 0.82745} & {\scriptsize 0.05389}
& {\scriptsize 0.03329}  & {\scriptsize 0.04047}  & {\scriptsize 0.05790}  \\
{\scriptsize $T_{2G}$} & {\scriptsize 3} & {\scriptsize 0,$\pm$3}        &
{\scriptsize 1}  & {\scriptsize 1.33333}
& {\scriptsize 1.99975} & {\scriptsize -1.63289} & {\scriptsize 0.85527} &
{\scriptsize 0.87308} & {\scriptsize -0.09762}
& {\scriptsize -0.08122} & {\scriptsize -0.12181} & {\scriptsize 0.09947} \\
{\scriptsize $G_G$}    & {\scriptsize 3} & {\scriptsize $\pm$1,$\pm$2}   &
{\scriptsize 1}  & {\scriptsize 0.50000}
& {\scriptsize 0.13364} & {\scriptsize -0.23822} & {\scriptsize 0.87251} &
{\scriptsize 0.87134} & {\scriptsize -0.16449}
& {\scriptsize -0.00908} & {\scriptsize -0.00303} & {\scriptsize 0.00416}
\\ \hline
{\scriptsize $T_{1U}$} & {\scriptsize 1} & {\scriptsize 0,$\pm$1}        &
{\scriptsize -1} & {\scriptsize 1.54204}
& {\scriptsize 1.50000} & {\scriptsize 1.50777}  & {\scriptsize 0.85363} &
{\scriptsize 0.84560} & {\scriptsize 0.19439}
& {\scriptsize 0.16148}  & {\scriptsize 0.11124}  & {\scriptsize 0.13512}  \\
{\scriptsize $H_U$}    & {\scriptsize 2} & {\scriptsize 0,$\pm$1,$\pm$2} &
{\scriptsize -1} & {\scriptsize 0.45796}
& {\scriptsize 0.50000} & {\scriptsize -0.08977} & {\scriptsize 0.87070} &
{\scriptsize 0.87873} & {\scriptsize 0.08662}
& {\scriptsize -0.01309} & {\scriptsize -0.02673} & {\scriptsize 0.00445}  \\
{\scriptsize $T_{2U}$} & {\scriptsize 3} & {\scriptsize 0,$\pm$3}        &
{\scriptsize -1} & {\scriptsize 0.63864}
& {\scriptsize 1.00000} & {\scriptsize 0.21489}  & {\scriptsize 0.86786} &
{\scriptsize 0.91547} & {\scriptsize -0.18339}
& {\scriptsize 0.02377}  & {\scriptsize -0.01409} & {\scriptsize 0.01251}  \\
{\scriptsize $G_U$}    & {\scriptsize 3} & {\scriptsize $\pm$1,$\pm$2}   &
{\scriptsize -1} & {\scriptsize 1.54204}
& {\scriptsize 1.25000} & {\scriptsize -1.17978} & {\scriptsize 0.85363} &
{\scriptsize 0.81391} & {\scriptsize -0.11652}
& {\scriptsize -0.06371} & {\scriptsize -0.07559} & {\scriptsize 0.07082}
\\ 
\end{tabular}
\end{table}


\begin{references}
\bibitem{IN} In the standard spectroscopic notation an additional
             upper index 2S+1, which gives
             the multiplicity of the states, is necessary for a complete 
             characterization. However, in the considered case of singlet 
             states ($S=0$) the multiplicity is fixed to the value 1 and, 
             hence, may be omitted.
\end{references}
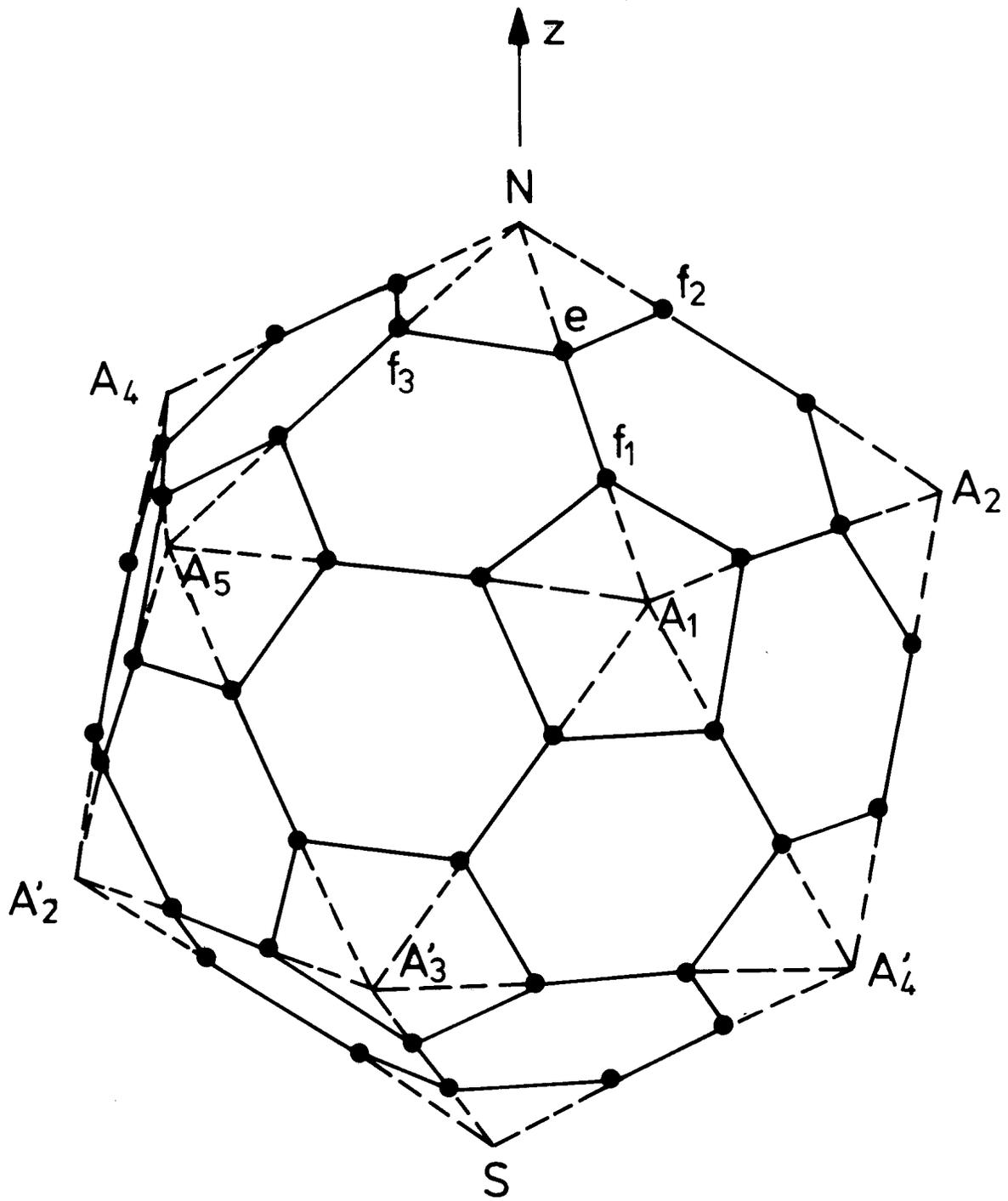
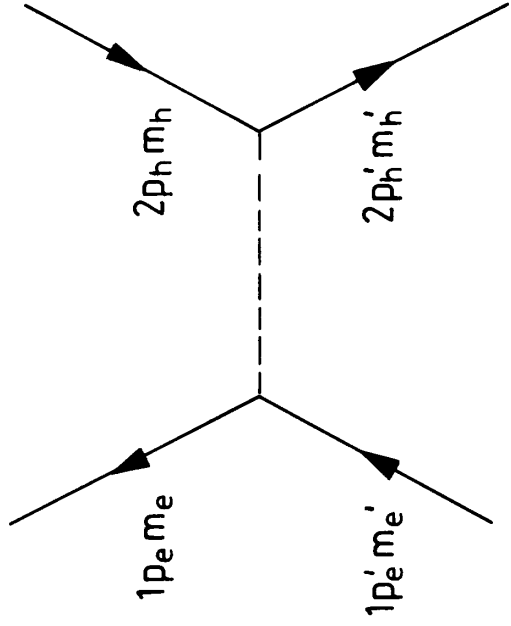
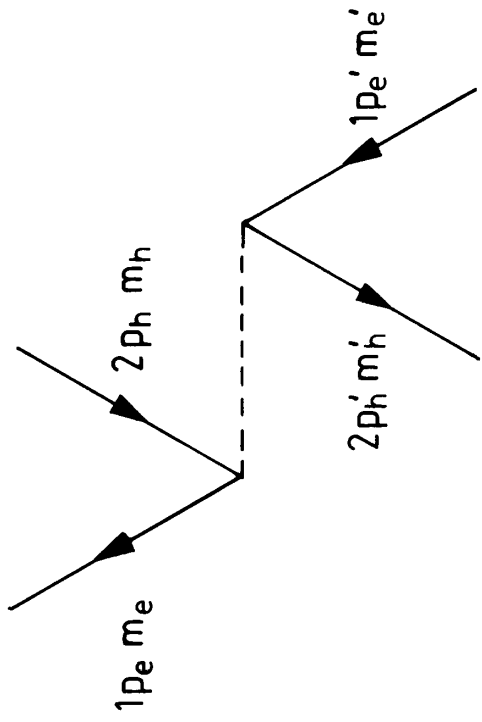
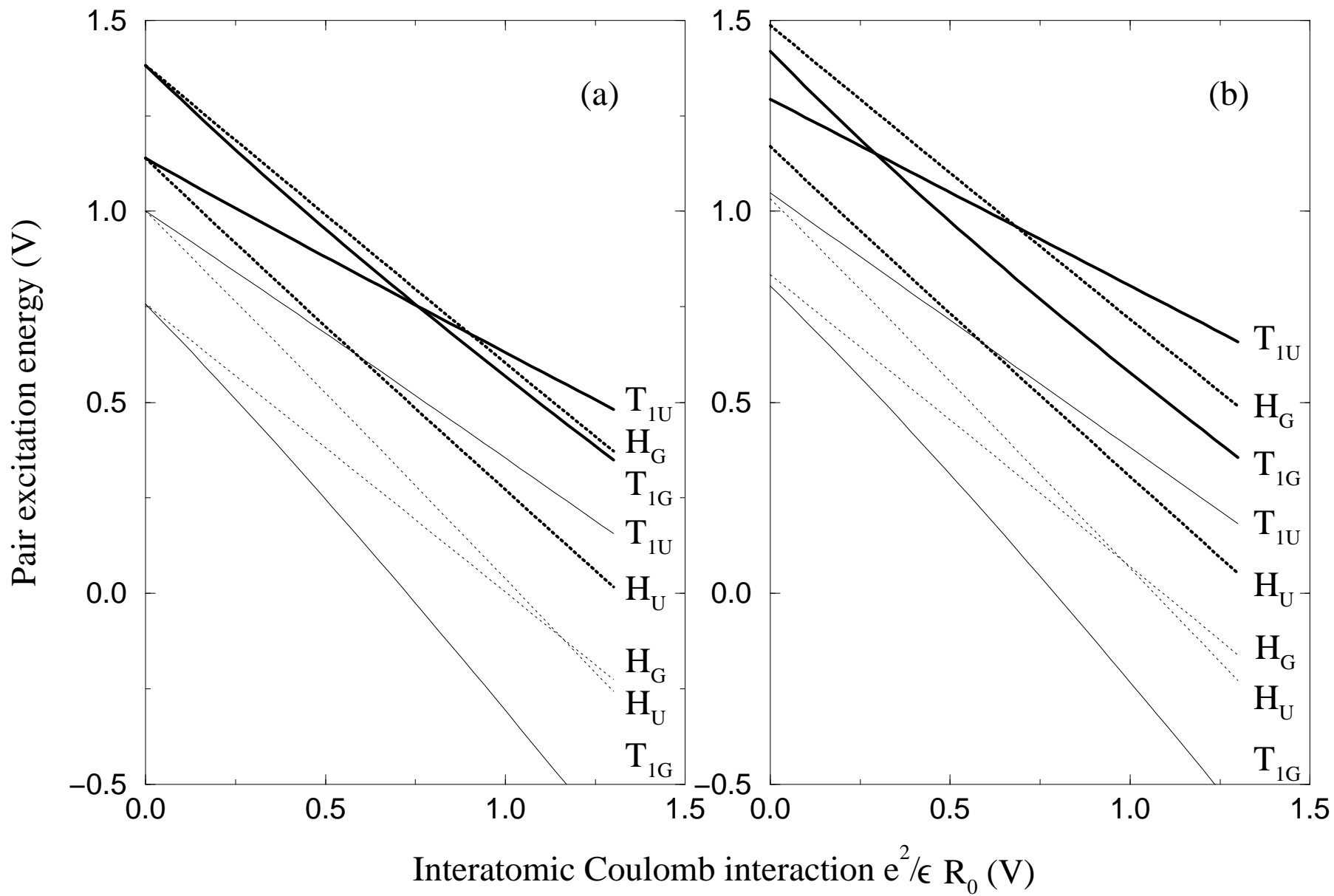
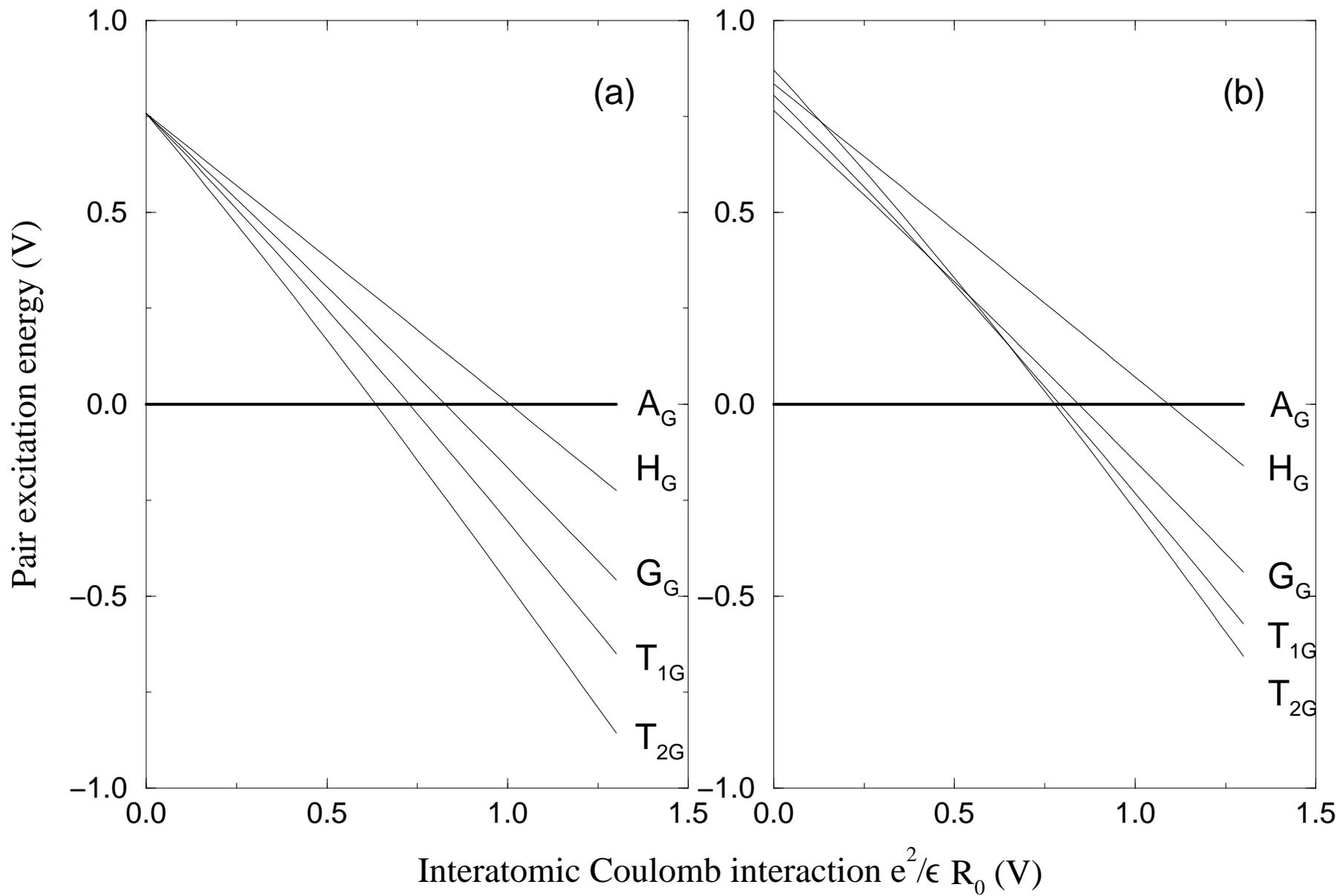
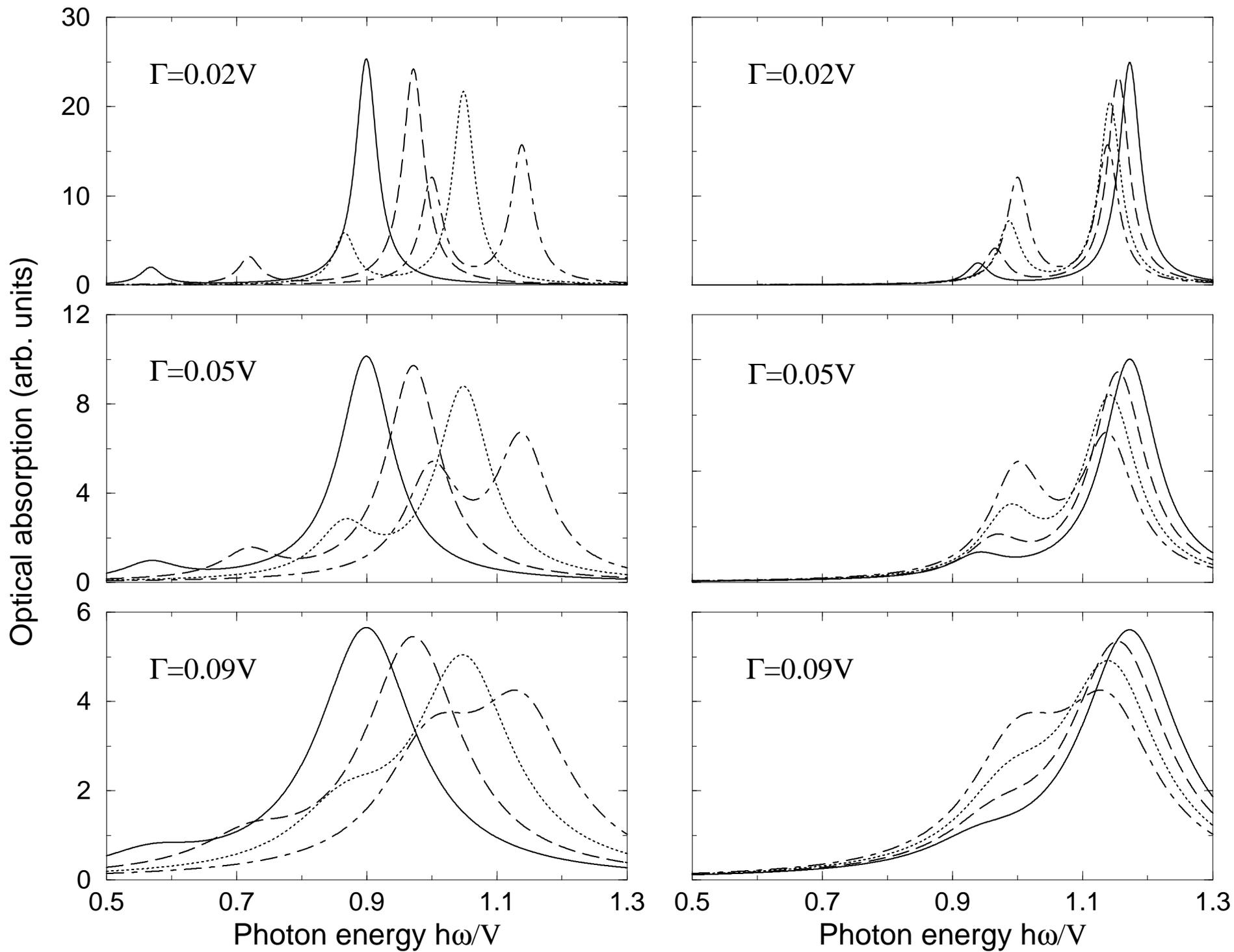
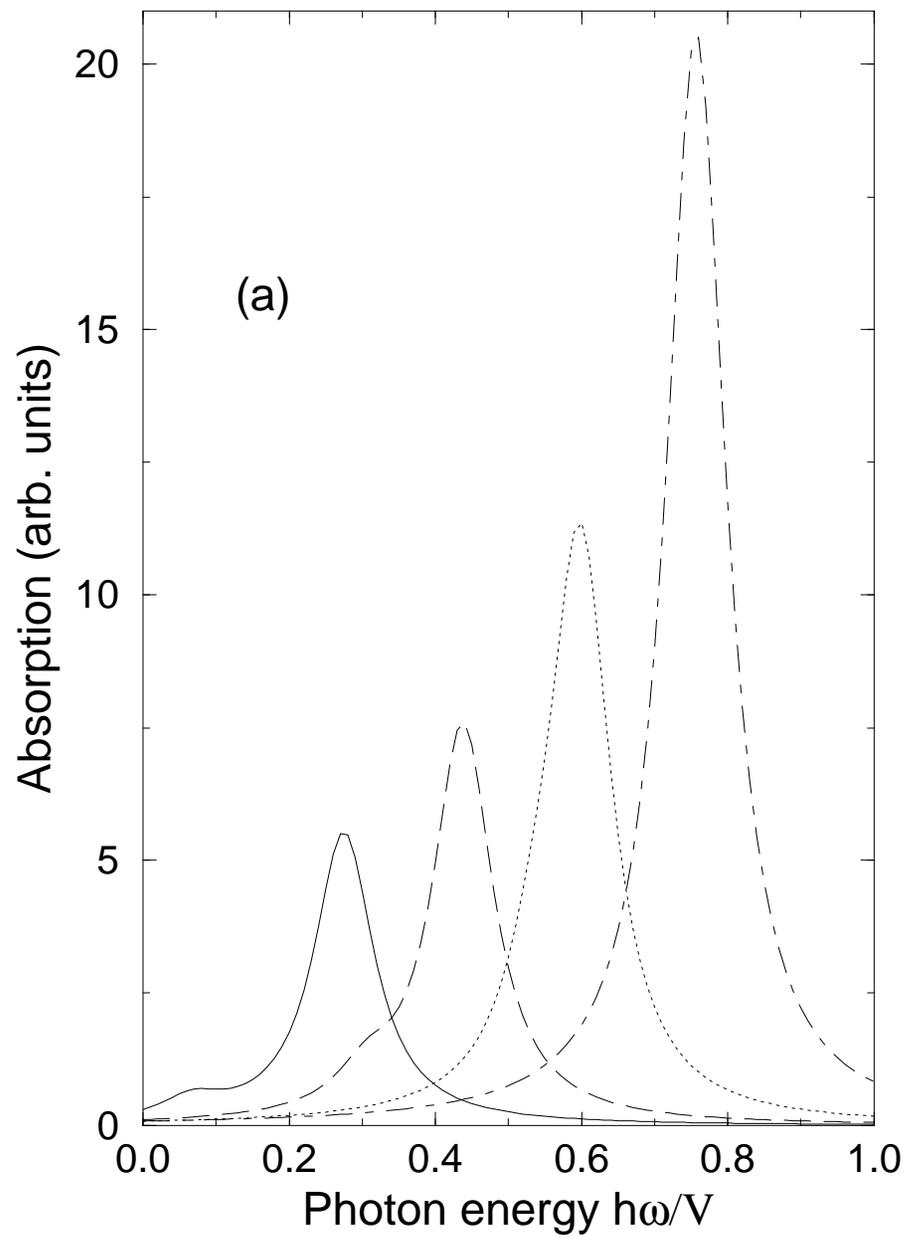
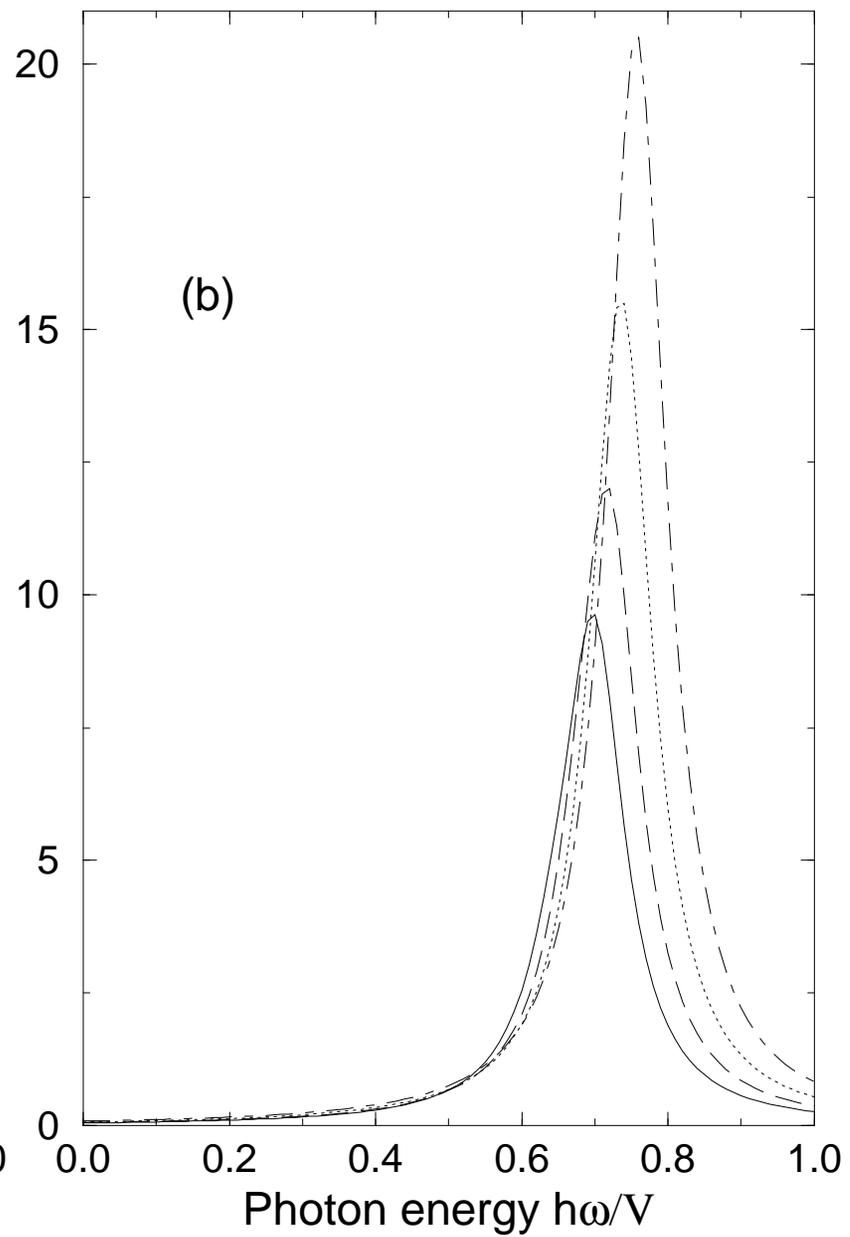
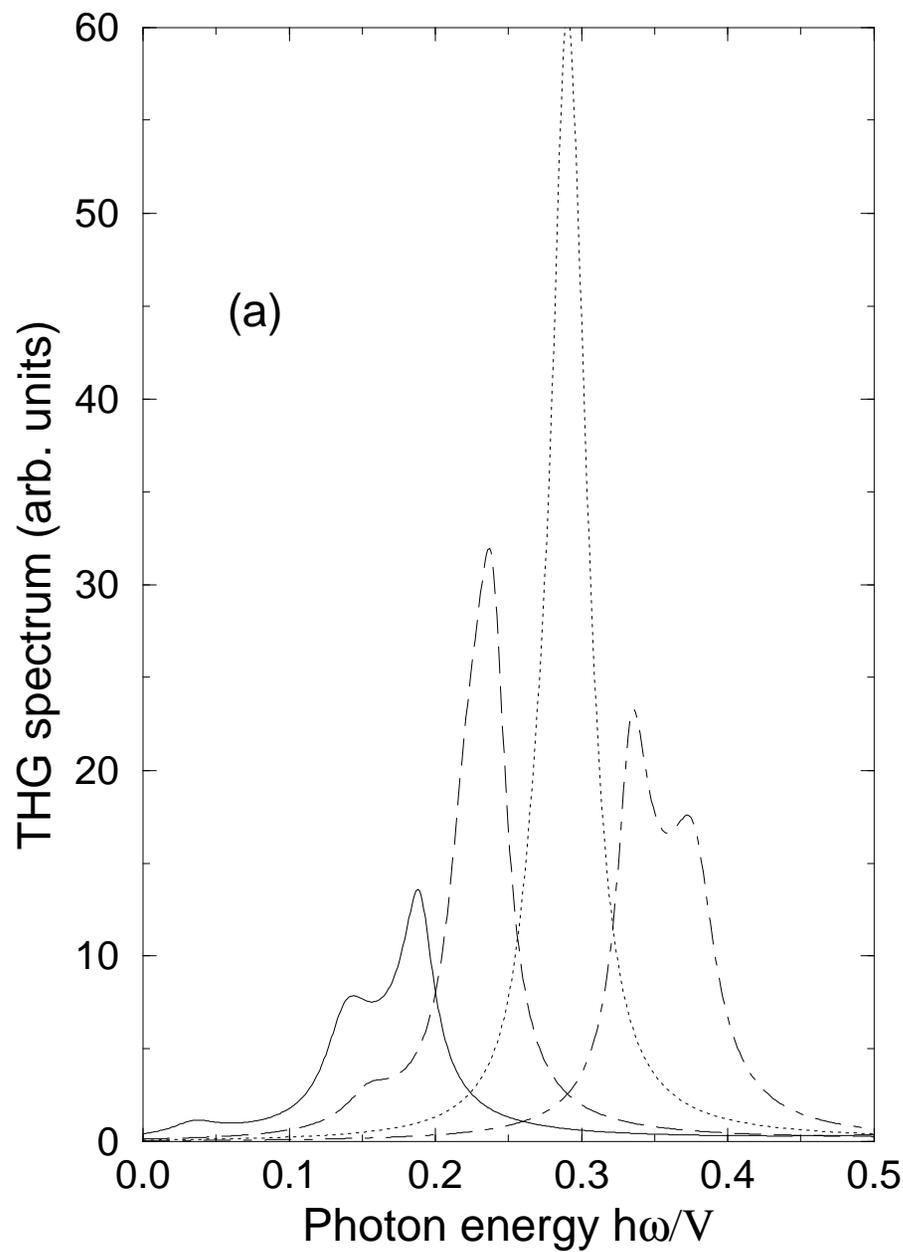
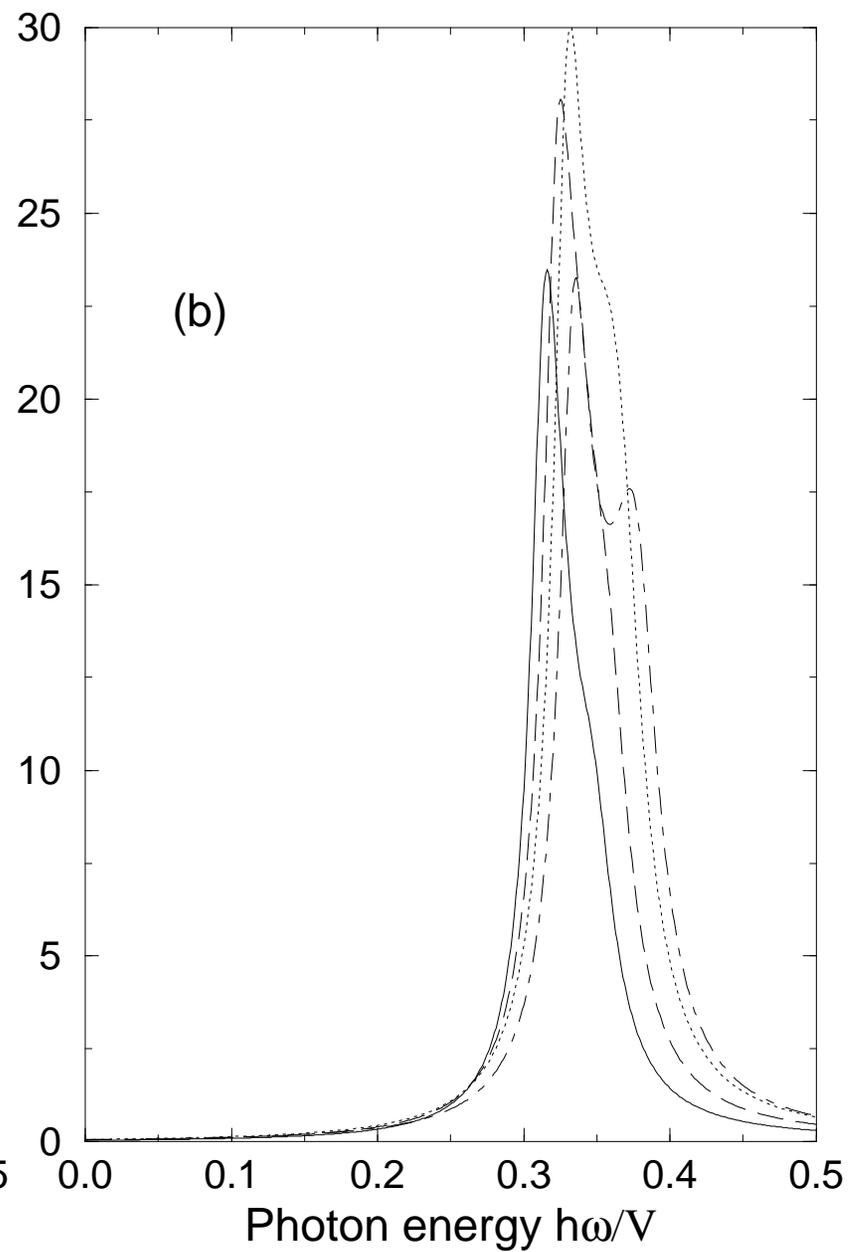
\end{document}